\documentclass{JHEP3}

\pdfoutput=1

\usepackage{graphicx}
\usepackage{amssymb}
\usepackage{slashed}
\usepackage{amsmath}
\usepackage{url}

\newcommand{\be}{\begin{equation}}
\newcommand{\ee}{\end{equation}}
\newcommand{\Eq}[1]{Eq.~(\ref{#1})}
\newcommand{\Sec}[1]{Sec.~\ref{#1}}
\newcommand{\Fig}[1]{Fig.~\ref{#1}}
\newcommand{\Tab}[1]{Table~\ref{#1}}
\newcommand{\Ref}[1]{Ref.~\cite{#1}}

\title{Jet Trimming}

\author{David Krohn$^a$, Jesse Thaler$^{b,c}$, and Lian-Tao Wang$^a$ \\ 
$^a$Department of Physics, Princeton University, Princeton, NJ 08540  \\ 
$^b$Berkeley Center for Theoretical Physics, University of California, \\
\ Berkeley, CA 94720, U.S.A.\\
$^c$Theoretical Physics Group, Lawrence Berkeley, National Laboratory,\\
\ Berkeley, CA 94720, U.S.A.\\
\ E-mail:  \email{dkrohn@princeton.edu,jthaler@jthaler.net,lianwang@princeton.edu} }
\abstract{Initial state radiation, multiple interactions, and
event pileup can contaminate jets and degrade event reconstruction.
Here we introduce a procedure, jet trimming, designed to mitigate these sources of contamination in 
jets initiated by light partons.  This procedure is complimentary to existing methods developed for boosted heavy particles.
We find that jet trimming can achieve significant improvements in event reconstruction, especially at high energy/luminosity hadron colliders 
like the LHC.}


\begin{document}


\section{Introduction}

Jets are collections of hadronic four-momenta used to approximate the kinematics of short distance scattering events.   Since the high-energy frontier is explored by hadron colliders with color-rich final states, jets are a necessary tool to better understand the physics of the standard model and probe whatever lies beyond it.  To assemble jets one must make use of jet algorithms---well-defined procedures for collecting 
detector tracks and calorimeter cells into jet four-momenta.  Many such algorithms exist, with each exhibiting a different clustering behavior.\footnote{For comprehensive reviews and relevant references see Refs.~\cite{Ellis:2007ib,Salam:2009jx}.}   Though the choice of jet algorithm introduces some level of ambiguity in any jet-based measurement, this is still acceptable, as any infrared/collinear-safe jet algorithm will yield results that can be compared to theoretical calculations.

In general, the optimal jet algorithm for an analysis is the one which most closely reconstructs the hard scattering process.   The closer the reconstruction is to the true scattering, the greater the signal significance.\footnote{In principle, the choice of jet algorithm could also help control reducible backgrounds.}  Now, if the final states observed in a detector only arose from the products of a hard scattering, and if the jets were well-separated from each other and from the beamline, then the precise jet definition used would not matter very much.  In that idealized scenario, the jets would be accurately reconstructed by any jet algorithm, as long as the algorithm clustered most of the hadrons arising from final state radiation (FSR).

In reality, however, a detector records more than just the final states from a hard scattering event.  The incoming states 
will typically radiate before scattering, leading to copious initial state radiation (ISR).  ÊIn addition, multiple 
parton interactions (MI) and event pileup will further contaminate the final state.\footnote{A hard scattering event takes place between the partons of two colliding hadrons.  Further interactions between those hadrons are called multiple interactions, while interactions between other hadrons in the colliding bunches are called pileup.}  This is an especially  prominent effect at the Large Hadron Collider (LHC) because of its high energy and luminosity. The net effect is that hadrons from ISR/MI/pileup are spatially overlapped with hadrons from FSR, complicating the jet finding procedure.  Thus, there is an inevitable tradeoff.  On the one hand, we would like a jet algorithm to form jets large enough to cluster all of the hard scattering decay products and account for wide angle FSR emissions.  On the other hand, we are constrained in how large our jets can become by inevitable contamination from hadrons unassociated with the hard scattering.

This conflict between missing radiation and contamination is usually resolved through a judicious choice of the jet size parameter (usually the jet radius $R$).  One can either fix the jet radius at an optimal value, or employ an algorithm designed to choose the optimal size on a jet-by-jet basis (e.g. the VR algorithm~\cite{Krohn:2009zg}).  It is possible to go a step further and statistically account for the sources of contamination by assuming a diffuse distribution and subtracting off a fixed contribution to each jet proportional to its area~\cite{Cacciari:2008gn}. However, one can take a more aggressive approach by actively working to identify and remove the radiation contaminating each jet.  The basic idea behind such an approach stems from the observation that there is usually only one hard scattering per event; all other sources of radiation (ISR/MI/pileup)
are likely to be much softer.  By going inside a jet and removing soft radiation (through a modification of the sequential clustering procedure
or through the use of subjets), reconstruction can be improved.

This idea of hierarchical radiation and its potential use in cleaning up contaminated jets has gained acceptance
in the jet community.  In the past, most studies focused on boosted hadronically decaying  
particles like the $W/Z$~\cite{Seymour:1993mx,Butterworth:2002tt}, Higgs~\cite{Butterworth:2008iy,Plehn:2009rk}, 
and top~\cite{Brooijmans,Kaplan:2008ie,Ellis:2009su, Ellis:2009me},\footnote{See Refs.~\cite{Butterworth:2007ke,Butterworth:2009qa} for some examples in supersymmetric processes.} where the procedure is optimized toward improving the jet mass resolution.  The only mention that we are aware of 
for using such a technique outside of heavy object reconstruction is \Ref{Cacciari:2008gd}, in which it was observed that applying the same 
procedure useful in reconstructing a boosted Higgs could also help reconstruct jets from light partons.

In this paper, we present procedures specifically designed to improve the reconstruction of ordinary QCD jets arising from the showering and fragmentation of nearly massless partons (i.e.\ light quarks and gluons).   To distinguish this from prior work on boosted heavy particles (such as jet filtering \cite{Butterworth:2008iy} and jet pruning \cite{Ellis:2009su}), we will call our procedures \emph{jet trimming}.  In the next section, we will further discuss the contamination of jets and try to quantify its effects.    In \Sec{sec:thealgorithm}, we will introduce jet trimming algorithms and discuss different versions of these applicable to final states in various kinematical regimes.  In \Sec{sec:results}, we will present the results of our algorithms and compare them both with the untrimmed results and with earlier cleaning techniques.  We will see that by using algorithms  specifically designed for light parton jets we can achieve a substantial gain, beyond the improvements seen through applying the techniques developed for boosted heavy particles.  \Sec{sec:conclusions} contains our conclusions.

\section{Trimming QCD Jets}
\label{sec:benefits}

As discussed in the introduction, jet reconstruction always presents a trade off between capturing all of the radiation associated with a hard scattering while at the same time minimizing the contamination from other hadrons  present in an event.  Before we discuss this, let us first introduce some notation and provide some details about our study.  

Throughout this paper, we will refer the typical size of a jet in terms of its characteristic radius $R$ using distances defined  on the (rapidity $y$, azimuth $\phi$) plane:  $\Delta R = \sqrt{ (\Delta y)^2 + (\Delta \phi)^2}$ .  When referring to generic fixed-radius jets and their size ($R_0$), we are implicitly using the anti-$k_T$ algorithm~\cite{Cacciari:2008gp} for jet reconstruction, as this reasonably approximates the behavior of  an ideal cone algorithm (for a discussion on the behavior of other algorithms in reconstruction see Ref.~\cite{Cacciari:2008gd}).  To generate our  Monte Carlo events samples we use \texttt{Pythia 6.4.21}~\cite{Sjostrand:2006za} with the default `Tune-A'~\cite{Field:2002vt,Field:2005sa} settings
and assume a $14~{\rm TeV}$ LHC.   Our jets are clustered using  \texttt{FastJet 2.4.0}~\cite{Cacciari:Fastjet, Cacciari:2005hq}.
While the discussion here in \Sec{sec:benefits} will not account for the effects of pileup (so as to demonstrate the irreducible, significant effects of ISR/MI contamination), we will factor in the effects of pileup for our results in \Sec{sec:results}, assuming a relatively modest  luminosity per bunch-bunch crossing of $0.05~{\rm mb}^{-1}$.  To approximate the effects of a real detector, we always group final state partons/hadrons  into $\delta \eta \times \delta \phi = 0.1 \times 0.1$ calorimeter cells between $-3 < \eta < 3$, and assign the cells massless four-momenta  based on the calorimeter energy.

Finally, we note that while most aspects of particle collisions calculated in Monte Carlo programs rest on firm bases from fundamental physics, the effects of hadronizaton are only understood through phenomenological models.\footnote{Hadronizaton is modeled in Pythia using the Lund model~\cite{Andersson:1998tv}, which has been successful in reproducing collider data~\cite{Barate:1996fi}.}  This might seem to be cause for concern, as our results will to some extent reflect the effects of hadronization, but we expect these dependencies to be small, altering perturbatively calculated jet/subjet momenta by ${\cal O}(\Lambda_{\rm QCD})$.  While we will operate under this assumption for the rest of the article, the validation of hadronization models will be an important task at the LHC.

\subsection{The Effects of Contamination}

\FIGURE[t]{
\includegraphics[scale=0.32]{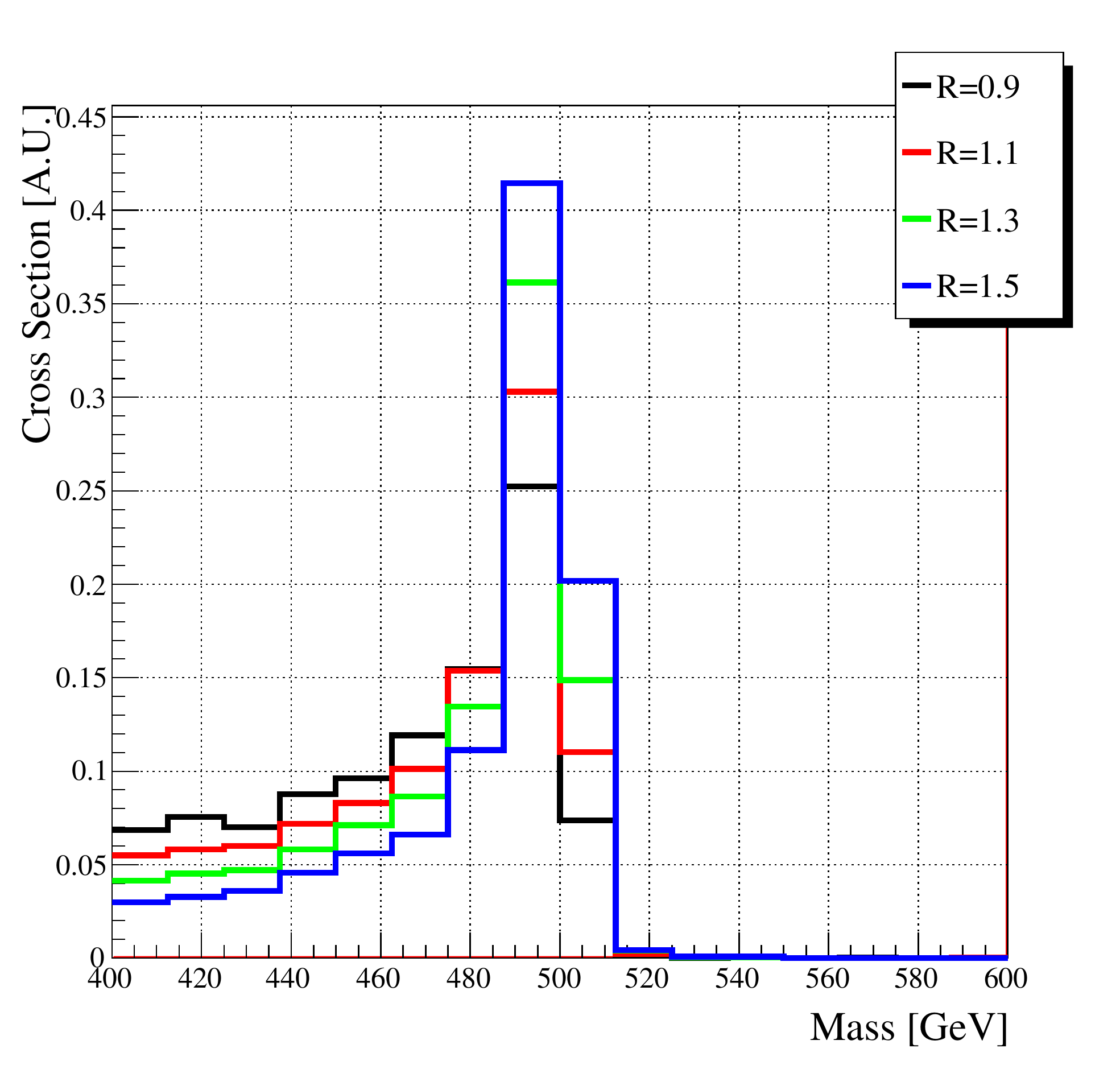}
\includegraphics[scale=0.32]{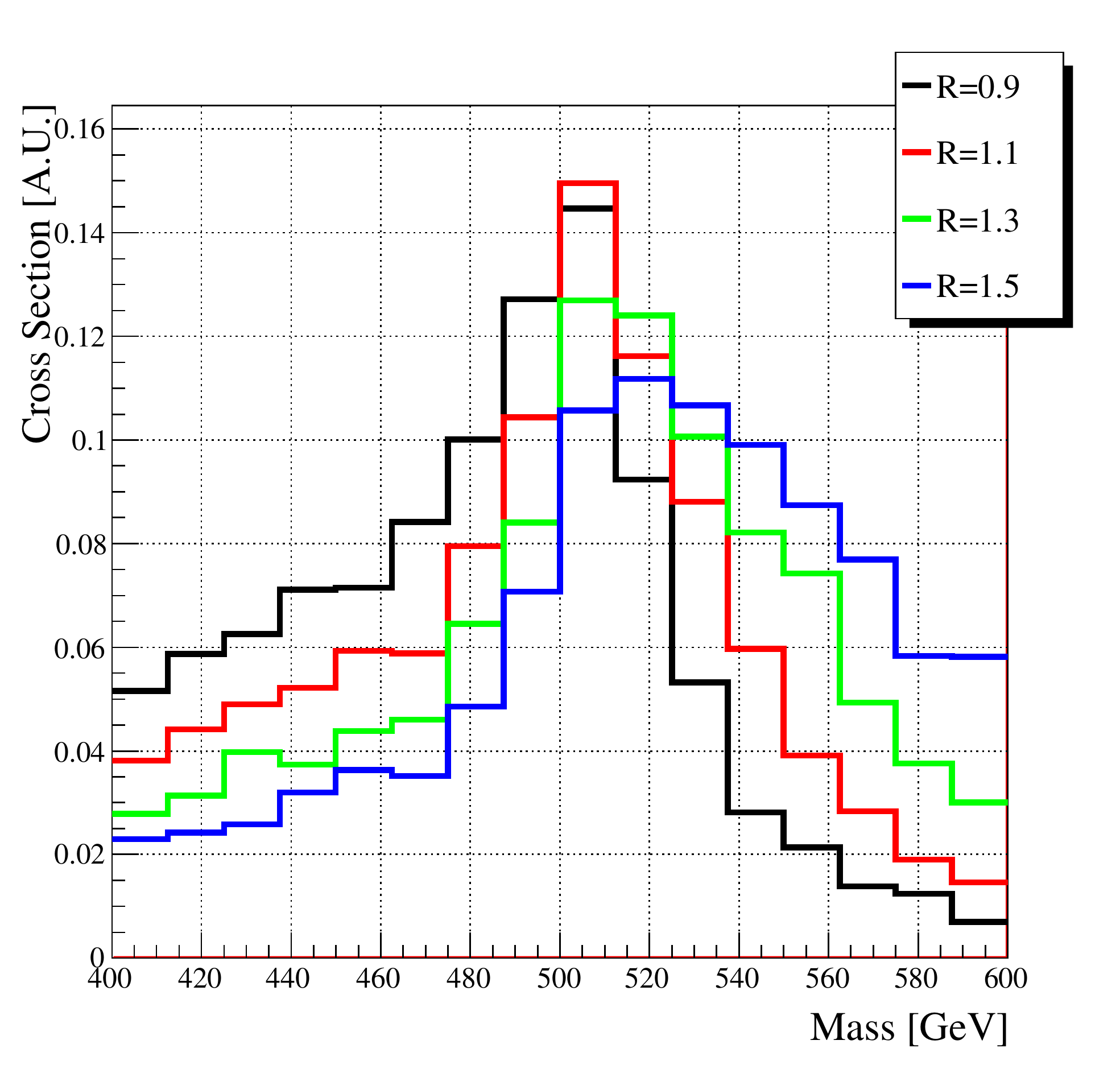}
\caption{Reconstructed $m_\phi=500~{\rm GeV}$ from $gg \rightarrow \phi \rightarrow gg$ dijet events with FSR only (left) and with the addition of ISR/MI (right).  In the absence of ISR/MI larger jet radii are preferred, while when ISR/MI are turned on a smaller radii must be used to balance the effects of contamination.}
\label{fig:RwithFSRonly}
\label{fig:RwithEverythingOn}
}

In absence of ISR/MI contamination, a large $R$ is desirable in the context of traditional jet clustering.  To see why, consider the process $gg\rightarrow \phi\rightarrow gg$ where $\phi$ is a new color octet scalar with a mass of $500~{\rm GeV}$ and a narrow width.\footnote{The $\phi$ couples to gluons via the  operator $\mathrm{Tr}(\phi G_{\mu \nu} G^{\mu \nu})$.  For comparison, we will also consider a different color octet scalar $\phi$ that couples to fermions via a Yukawa coupling $\bar{q} \phi q$.}  In a showering Monte Carlo program without hadronization, FSR is factorized from ISR/MI, so one can study the FSR in isolation.\footnote{With hadronization turned on, there are non-trivial correlations between FSR and ISR.}  On the left side of  \Fig{fig:RwithFSRonly},  we show the distribution of the reconstructed $\phi$ mass using only FSR for various values of the anti-$k_T$ jet radius $R_0$.  One sees that as $R_{0}$ increases, the reconstructed invariant mass distribution approaches the narrowly peaked distribution predicted from the hard scattering.

However, when one includes the effect of contamination, larger values of $R_{0}$ can yield poorer reconstruction, as seen from the right
 side of  \Fig{fig:RwithEverythingOn}.  Here, the jet radius that most closely matches the desired peak position is around $R_{\max} = 1.1$, considerably smaller than what one would want to use considering FSR alone.  From this one can see that an optimal jet algorithm would be one with a large overall jet radius that somehow avoids clustering in hadrons from ISR/MI (as well as pileup).

\FIGURE[t]{
\includegraphics[scale=0.32]{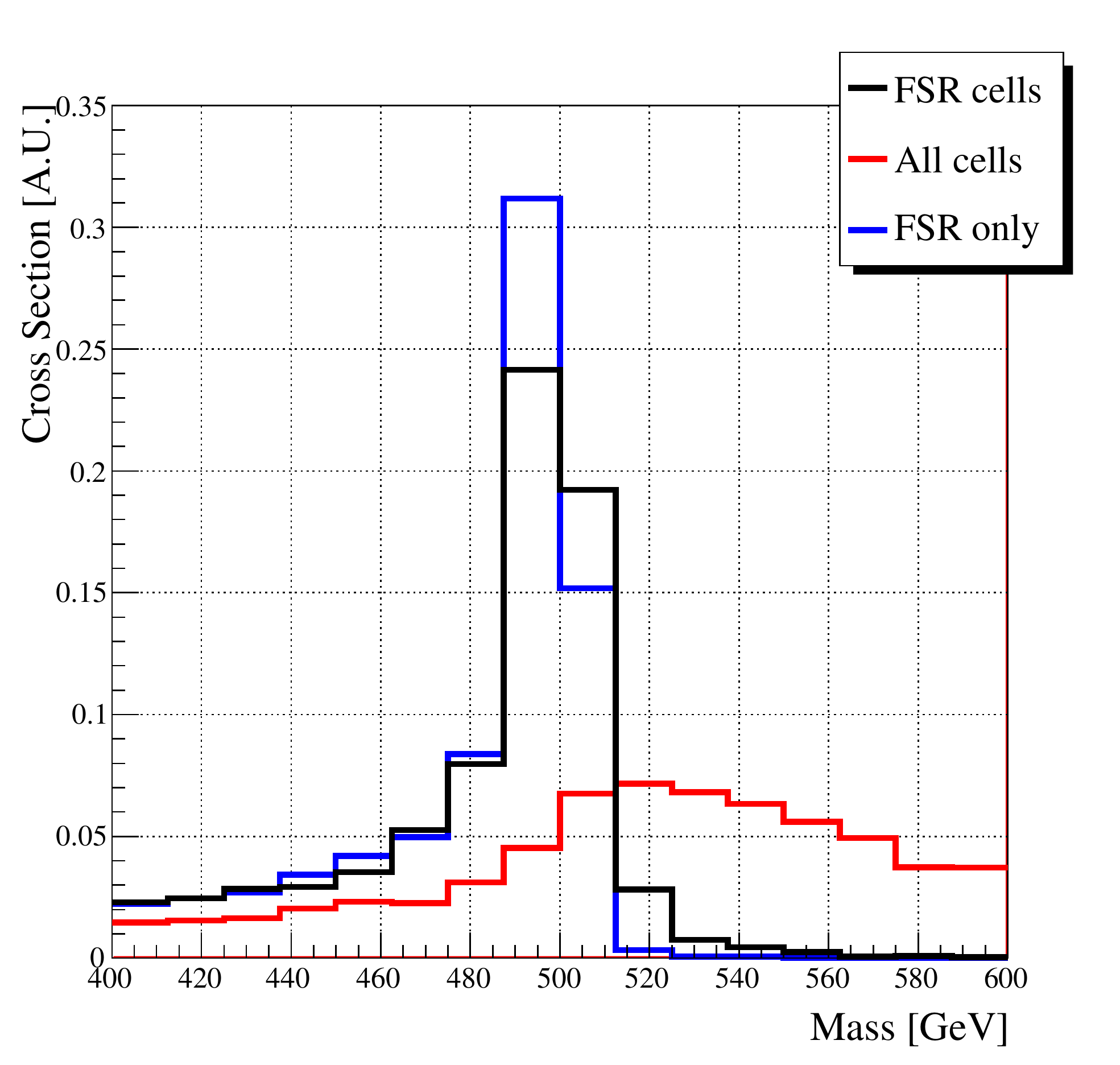}

\includegraphics[scale=0.32]{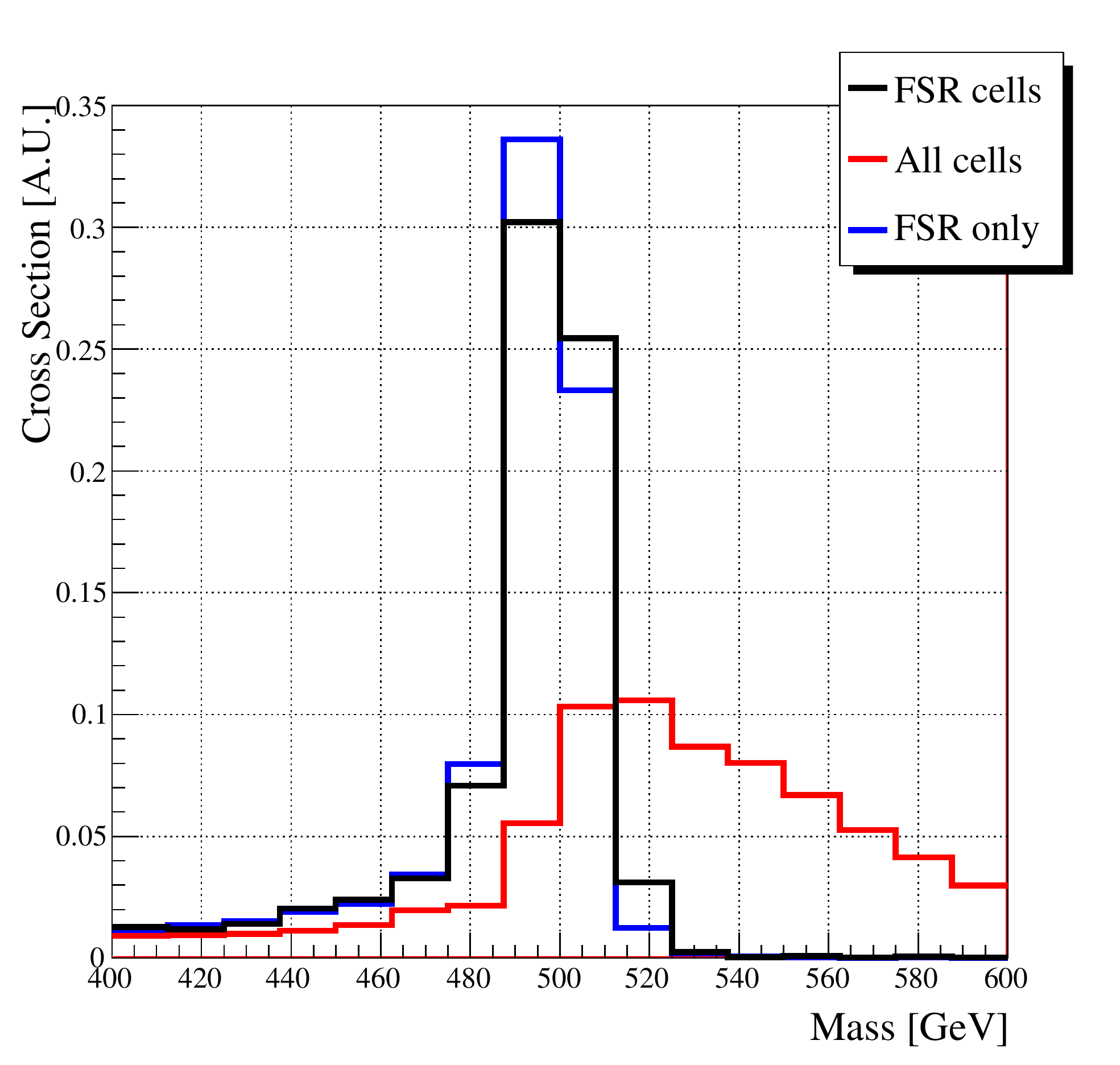}
\caption{Reconstructed $m_\phi=500~{\rm GeV}$ clustered with $R_0=1.5$ for $gg\rightarrow \phi \rightarrow gg$ (left) and $q\bar{q}\rightarrow \phi\rightarrow q\bar{q}$ (right).  The blue curve shows the reconstruction from a sample without ISR/MI.  
The red and black curves show data from a sample with ISR/MI, where all cells are clustered (red), and where only  those cells within $\Delta R=0.2$ of an cell containing more than $1~{\rm GeV}$ of FSR (black). The similarity between the black and blue peaks demonstrates that considerable gains in reconstruction are possible despite the irreducible overlap in radiation.}
\label{fig:RwithFSRdefinedCells}
}

\TABLE[t]{
\parbox{\textwidth}{
\begin{center}
\begin{tabular} {|c|c|c|c|c|c|c|}
\hline
\bf	& Improvement&$R_0$ &$\Gamma$ [GeV] & $M$ [GeV]\\
\hline
&\multicolumn{4}{|c|} {$gg \rightarrow\phi\rightarrow gg$} \\
\hline
All cells& -&  1.2& 69& 518 \\
FSR cells &  $309\%$ & 1.5 & 15 &501\\
  \hline
 &\multicolumn{4}{|c|} {$q\bar{q} \rightarrow\phi\rightarrow q\bar{q}$} \\
 \hline
 All cells& -&  0.8& 31& 505 \\
FSR cells &  $189\%$ & 1.5 & 11 &501\\
\hline
\end{tabular}
\end{center}
\caption{\label{tab:inPrincipleBest} Improvement in the resonance reconstruction measure $\Delta$ presented in \Sec{sec:results} in going from 
standard clustering ({\it All cells}) to an idealized situation where we only cluster  those cells within $\Delta R=0.2$ of an cell containing more than $1~{\rm GeV}$ of FSR ({\it FSR cells}).  Here $m_\phi=500~{\rm GeV}$.  The definitions of $\Gamma$ and $M$ appear in \Eq{eq:sig_dist}.  Because of the larger color charge of gluons compared to quarks, there is more radiation in the $gg \rightarrow \phi \rightarrow gg$ case compared to the $q \bar{q} \rightarrow \phi \rightarrow q \bar{q}$ case, so the potential improvement is correspondingly larger.}
}}

Now, there is always a minimum spatial overlap between FSR and ISR/MI from the fact that the two sources of hadrons could end up nearby in the detector.  Fortunately, this overlap is relatively small.  In \Fig{fig:RwithFSRdefinedCells} we present the $\phi$ mass reconstructed using $R_0=1.5$ where only those calorimeter cells within $\Delta R = 0.2$~\footnote{While at this point the choice of $\Delta R = 0.2$ is somewhat arbitrary, later in \Sec{sec:results} we will see that this is a reasonable subjet radius for use in trimming.} of  one containing at least 1 GeV of FSR were clustered, along with the distribution obtained without this restriction.  The restricted distribution is quite close to the one where only FSR was clustered, confirming the minimum spatial overlap.  By considering this sort of restriction to FSR-heavy cells, one can calculate the maximum 
  possible reconstruction improvement in going from ordinary cones to such an idealized jet algorithm.  This is shown in \Tab{tab:inPrincipleBest},  where the improvement is measured by the reconstruction measure $\Delta$ presented in \Sec{sec:results}. We see potential improvements of up to $3\times$ in reconstruction.  Of course, such an idealized jet algorithm cannot exist since no physical observable can distinguish between FSR and ISR/MI, but the room for improvement is compelling.

\FIGURE[t]{
\includegraphics[scale=0.32]{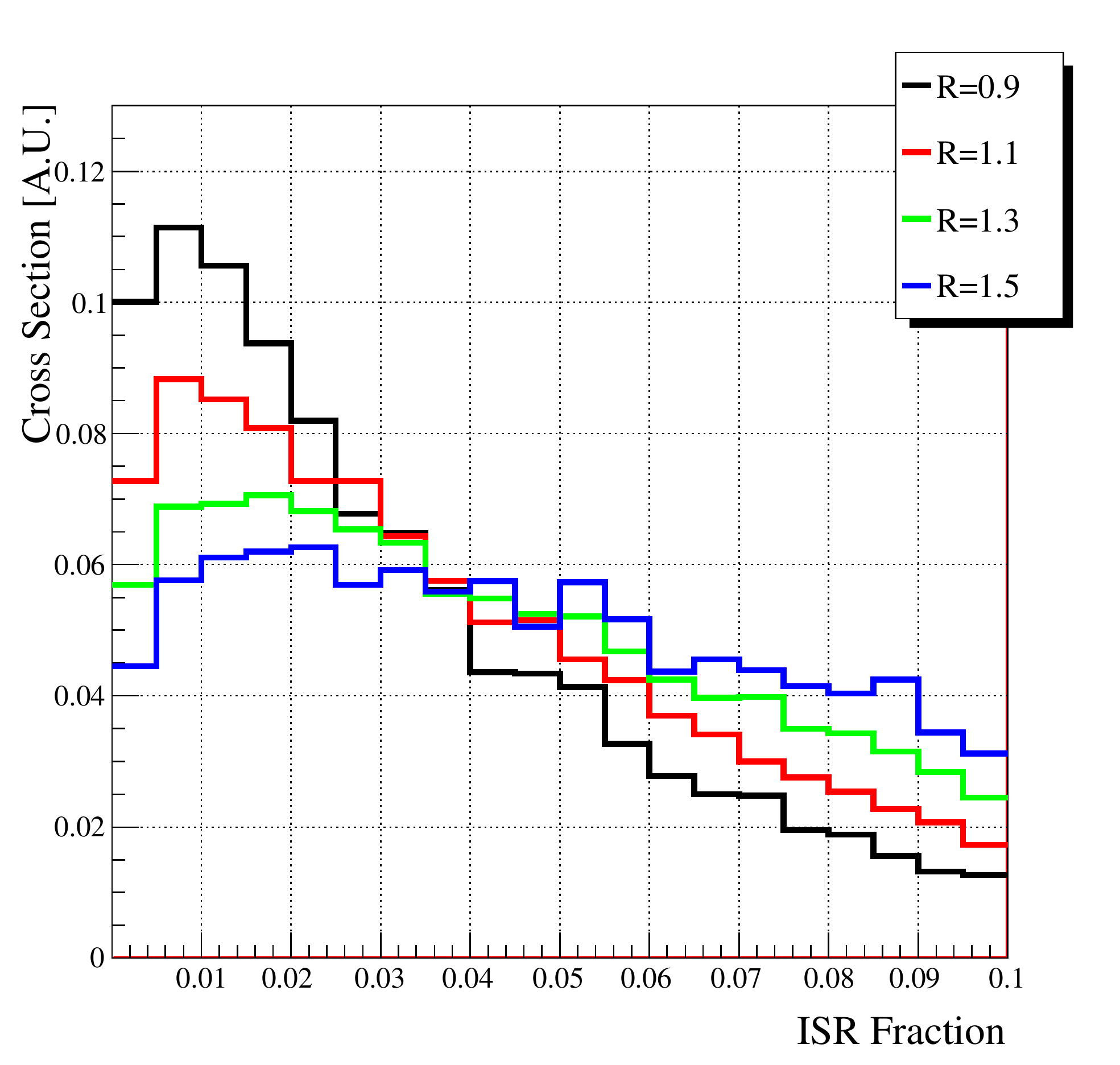}
\includegraphics[scale=0.32]{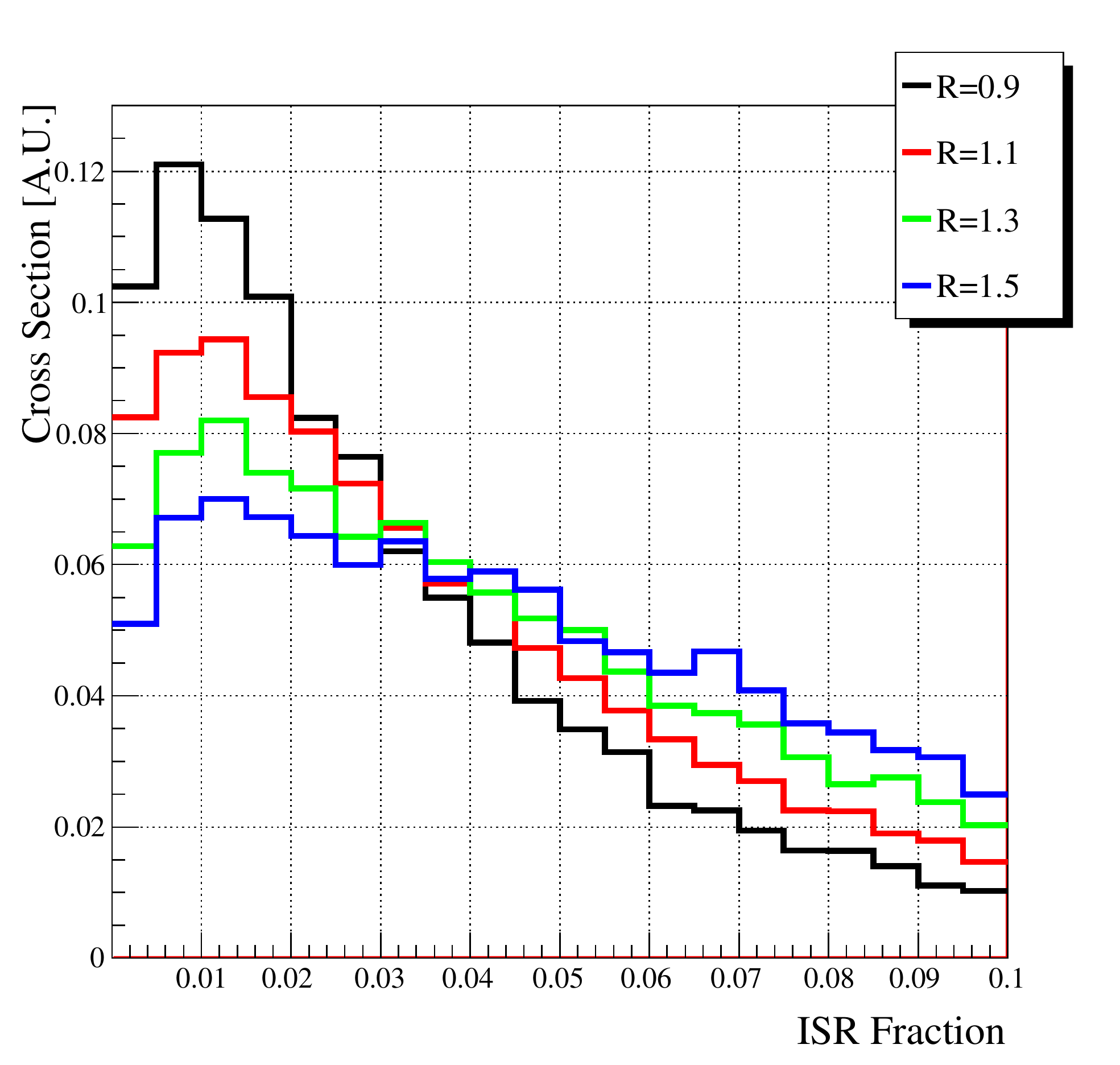}
\caption{Fraction of a jet's $p_T$ attributable to ISR/MI for $gg\rightarrow \phi\rightarrow gg$ (left) and $q\bar{q}\rightarrow \phi\rightarrow q\bar{q}$ (right).}
\label{fig:pTratio}
}

The goal our jet trimming algorithm is to approach this ideal reconstruction as closely as possible.  To do so, we need some kind of criteria to determine whether a given patch of the calorimeter is likely to contain substantial amounts of FSR.  In light of the observation that ISR/MI (as well as pileup) is usually soft compared to FSR, the simplest possible criteria we have is relative transverse momentum.  As shown in \Fig{fig:pTratio}, in a typical jet ISR/MI makes up only $\mathcal{O}(1-5\%)$  of the jet's $p_T$ (the contribution of pileup is a luminosity dependent question), and we saw earlier that there is minimal spatial overlap between contamination and FSR.  Therefore, sources of contamination can be mitigated by simply removing patches of soft calorimeter cells.

\subsection{QCD Jets vs. Boosted Objects}

While the general idea of removing soft calorimeter cells is straightforward, a number of details remain unspecified.  At minimum, one wants to consider patches of calorimeter cells by clustering them into subjets of radius $R_{\rm sub} > \delta_{\rm cal}=0.1$ to remove any sensitivity of the procedure to calorimeter segmentation.   Beyond that, one must specify how the subjets are to be formed, how large they will be, and what will serve as the criterion for softness.  As we will argue, by choosing jet trimming parameters in a way designed to enhance the reconstruction of light parton jets, we can increase reconstruction performance beyond the current techniques designed for boosted heavy particles~\cite {Seymour:1993mx,Butterworth:2002tt,Butterworth:2008iy,Plehn:2009rk,Brooijmans,Kaplan:2008ie,Ellis:2009su,Ellis:2009me,Butterworth:2007ke,Butterworth:2009qa}.

To see how one might go about choosing trimming parameters, consider first how they would be chosen to reconstruct the jet from a 
boosted heavy particle.  Usually such a particle decays immediately into two (e.g.\ the Higgs or $W$/$Z$) or three (e.g.\ the top) final states, each at the same characteristic $p_T$ scale (barring a matrix element conspiracy).   These states will shower into distinct hard patches in the jet (see the left panel of \Fig {fig:jet_comp}),  so one can hope to remove contamination from the system by simply associating a subjet 
to each hard final state and discarding everything else.  That is, one would discard all but the $N_{\rm cut}$ hardest subjets.  Whether or not a particular subjet  from a boosted heavy particle is considered soft depends upon where the subjet ranks in the subjet $p_T$ ordering and upon how many final state partons we expect in the decay.  For instance, if we are looking to reconstruct a Higgs in its decay $h\rightarrow b\bar{b}$ we would form subjets inside the initial jet and discard all but the hardest two.\footnote{Sometimes analyses allow for one subjet beyond the minimum number of tree level final states in order to capture the first emission~\cite{Butterworth:2008iy}, but the principle remains the same.}  In this context, the natural size of a subjet is also relatively clear; to treat each final state  of the decay equally (as we should, since they have comparable $p_T$s) we are limited to $R_{\rm sub} \lesssim R_0/2$ under the assumption that the initial jet was chosen to be just large enough to encompass the entire decay of the heavy particle.

 \FIGURE[t]{
\includegraphics[scale=0.32]{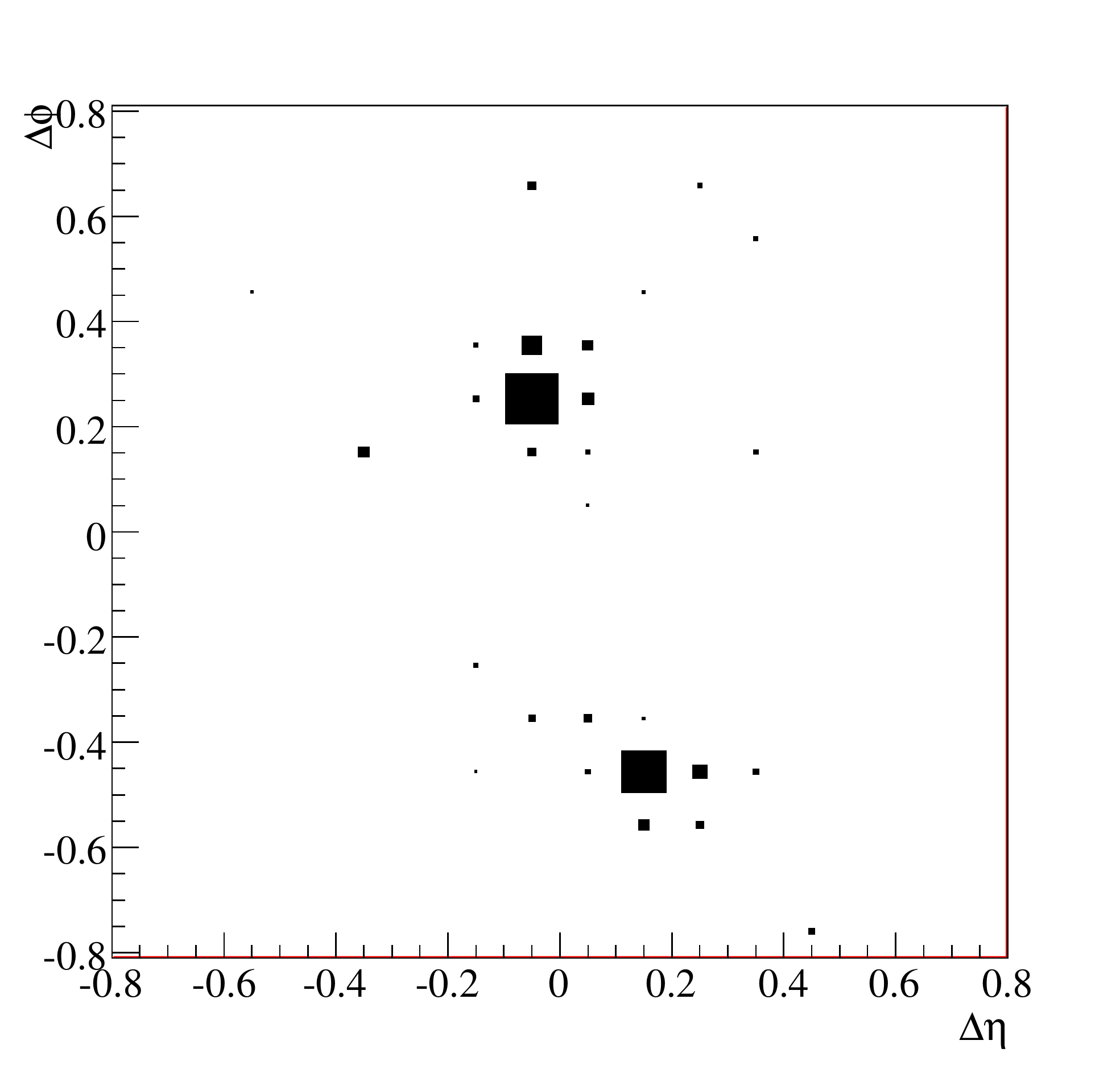}
\includegraphics[scale=0.32]{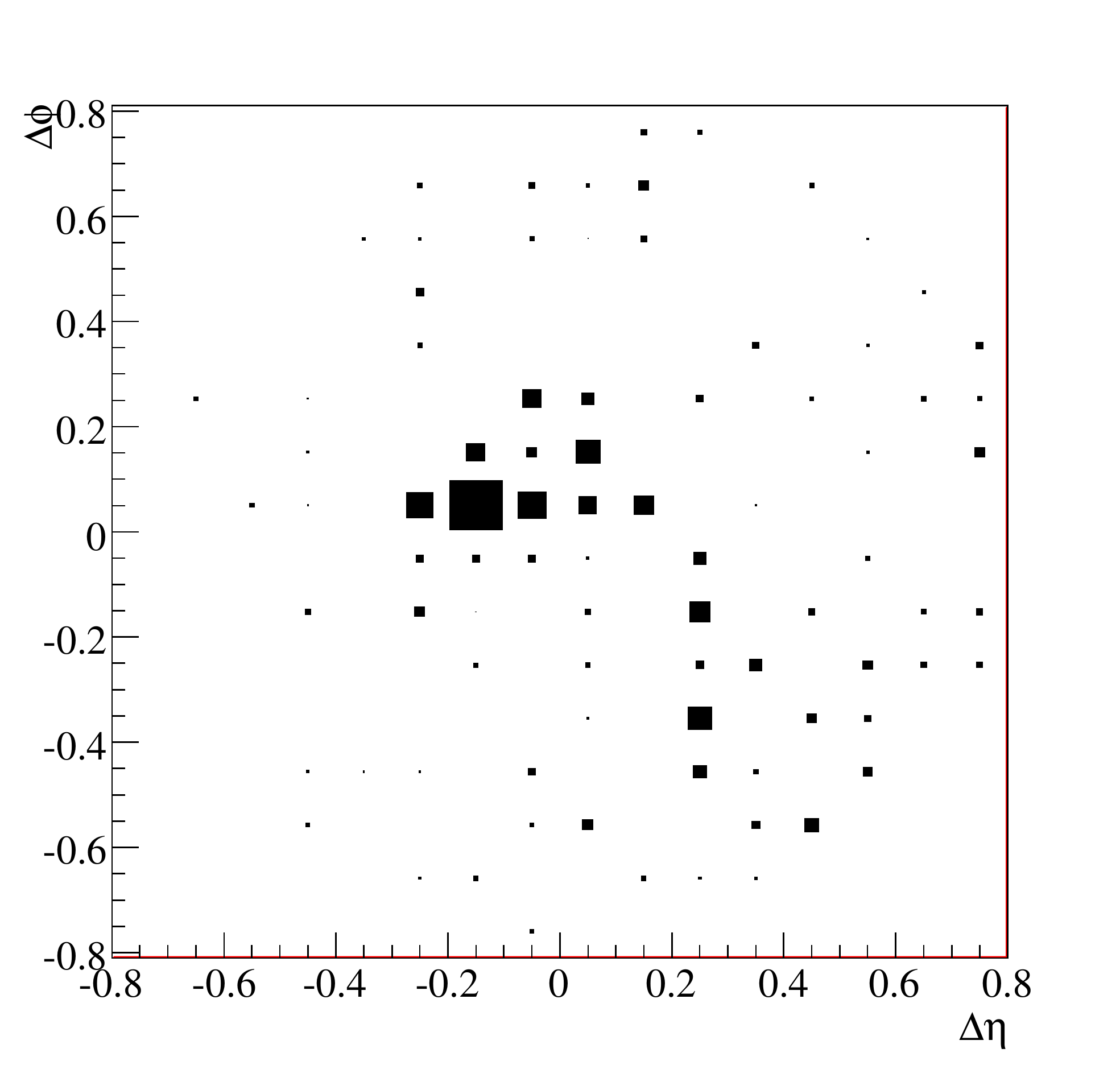}
\caption{Comparison of a jet formed from the decay of a boosted heavy particle (left) with one from the showering of light flavor/gluons (right).
Specifically, the left hand panel shows the jet formed from $h\rightarrow b\bar{b}$ while the right is a gluon jet.  The ($x$, $y$)-axes are ($y$, $\phi$)-distances as measured from the jet center and the area of each calorimeter
cell is proportional to its $p_T$.}
\label{fig:jet_comp}
}

The situation changes when we consider jets from light quarks or gluons (compare the two panels in \Fig {fig:jet_comp}).   The first difference is that there is only one hard final state at lowest order in $\alpha_s$.  Softness is therefore more naturally established directly via a cut on subjet $p_T$ rather than by restricting to a fixed number of subjets.  Later we will establish different subjet $p_T$ cuts for different kinematic regimes.  The second difference is that there is no natural size for the subjets as this depends upon the the $p_T$ cut for the subjets; a larger/smaller subjet size will necessitate a harder/softer subjet $p_T$ cut.  With these two differences in mind, we can now define our jet trimming procedure.

\section{Implementation}
\label{sec:thealgorithm}

In this section, we present an explicit algorithm implementing the jet trimming technique outlined above.\footnote{Our implementation
 is available as a plug-in to the \texttt{FastJet} package~\cite{Cacciari:Fastjet, Cacciari:2005hq}, which is available from the authors upon request.}    Our choice of algorithm is motivated primarily by simplicity and the ability to re-use existing jet finding procedures.  Many more 
 sophisticated choices could easily be imagined, but these are beyond the scope of the present work.
 
Since our jet trimming procedure will make use of well-known sequential recombination jet algorithms, we will briefly review 
how these work.   Recall that in a recursive jet algorithm one begins with an initial set of four-momenta (these could be tracks, calorimeter cells, etc.), 
assigning  every pair a  ``jet-jet distance 
measure'' $d_{ij}$ and every individual four-momenta a  ``jet-beam distance measure'' $d_{iB}$.  
The distance measures relevant for our study are:\footnote{For jet algorithm aficionados, we use ``VR'' to refer to the ``AKT-VR'' algorithm of Ref.~\cite{Krohn:2009zg}.}
\begin{align}
\text{anti-}k_T~\cite{Cacciari:2008gp}: & \qquad d_{ij} = \frac{1}{\max \left[p_{Ti}^{2}, p_{Tj}^{2} \right]} \frac{R_{ij}^2}{R_0^2} , \qquad d_{iB} = \frac{1}{p_{Ti}^{2}}, \\
\text{C/A}~\cite{Dokshitzer:1997in,Wobisch:1998wt}: & \qquad d_{ij} = \frac{R_{ij}^2}{R_0^2} , \qquad d_{iB} = 1, \\
k_T~\cite{Catani:1993hr,Ellis:1993tq}: & \qquad d_{ij} = \min \left[p_{Ti}^{2}, p_{Tj}^{2} \right] \frac{R_{ij}^2}{R_0^2} , \qquad d_{iB} = p_{Ti}^{2}\\
\text{VR}~\cite{Krohn:2009zg}: & \qquad d_{ij} = \frac{1}{\max \left[p_{Ti}^{2}, p_{Tj}^{2} \right]}{R_{ij}^2} , \qquad d_{iB} = \frac{\rho^2}{p_{Ti}^4}.
\end{align}
At each step in the clustering, the smallest entry in the set of all $d_{ij}$ and $d_{iB}$ is identified.  When a jet-jet distance is the smallest, 
the corresponding four-momenta are merged, while if a jet-beam distance is the smallest, then the associated four-momentum is  ``merged 
with the beam'' and set aside.  Here we will deal entirely with inclusive algorithms, where the recursion continues until all jets are merged with the beam, and the algorithm returns those merged jets whose $p_T$ is greater than some minimum value.\footnote{In an exclusive algorithm, the recursion  stops when a predetermined distance measure $d_{\rm cut}$ is reached, at which point the unmerged jets are returned.} 

\subsection{Jet Trimming}
\label{sec:trimalg}

The jet trimming procedure we advocate is an  ``outside-in'' algorithm, meaning that a seed jet determined through one jet finding method is reclustered using a subjet finding method.  Then a softness criteria is applied to the individual subjets to determine the final trimmed jet.  One could also imagine an ``inside-out'' algorithm, where small  subjets are found first, and clustering into a larger jet, again using some kind of softness criteria, but we will not explore that option here.

The proposed algorithm proceeds as follows:
\begin{enumerate}
\item Cluster all cells/tracks into jets using any clustering algorithm.  The resulting jets are called the seed jets.
\item Within each seed jet, recluster the constituents using a (possibly different) jet algorithm into subjets with a characteristic radius $R_{\rm sub}$
smaller than that of the seed jet.  
\item  Consider each subjet, and discard the contributions of
subjet $i$ to the associated seed jet if $p_{Ti} < f_{\rm cut} \cdot \Lambda_{\rm hard}$, where $f_{\rm cut}$ is a fixed dimensionless parameter, and $\Lambda_{\rm hard}$ is some hard scale chosen depending upon the kinematics of the event.
\item Assemble the remaining subjets into the trimmed jet.
\end{enumerate}
This procedure is illustrated in Figs.~\ref{fig:trimming_step_by_step_log} and \ref{fig:trimming_step_by_step_linear}.  The dimensionless parameter $f_{\rm cut}$ quantifies the expected $p_T$ scale hierarchy between FSR and ISR/MI/pileup.  In principle, this procedure could be iterated such that subjets that fail the softness criteria in one seed jet could be tested for inclusion in a different seed jet.  However, this is only relevant if the original jets were effectively overlapping, or if the removal of subjets substantially changes the position of the trimmed jets relative to the original seed jets.

\FIGURE[t]{
\includegraphics[scale=0.32]{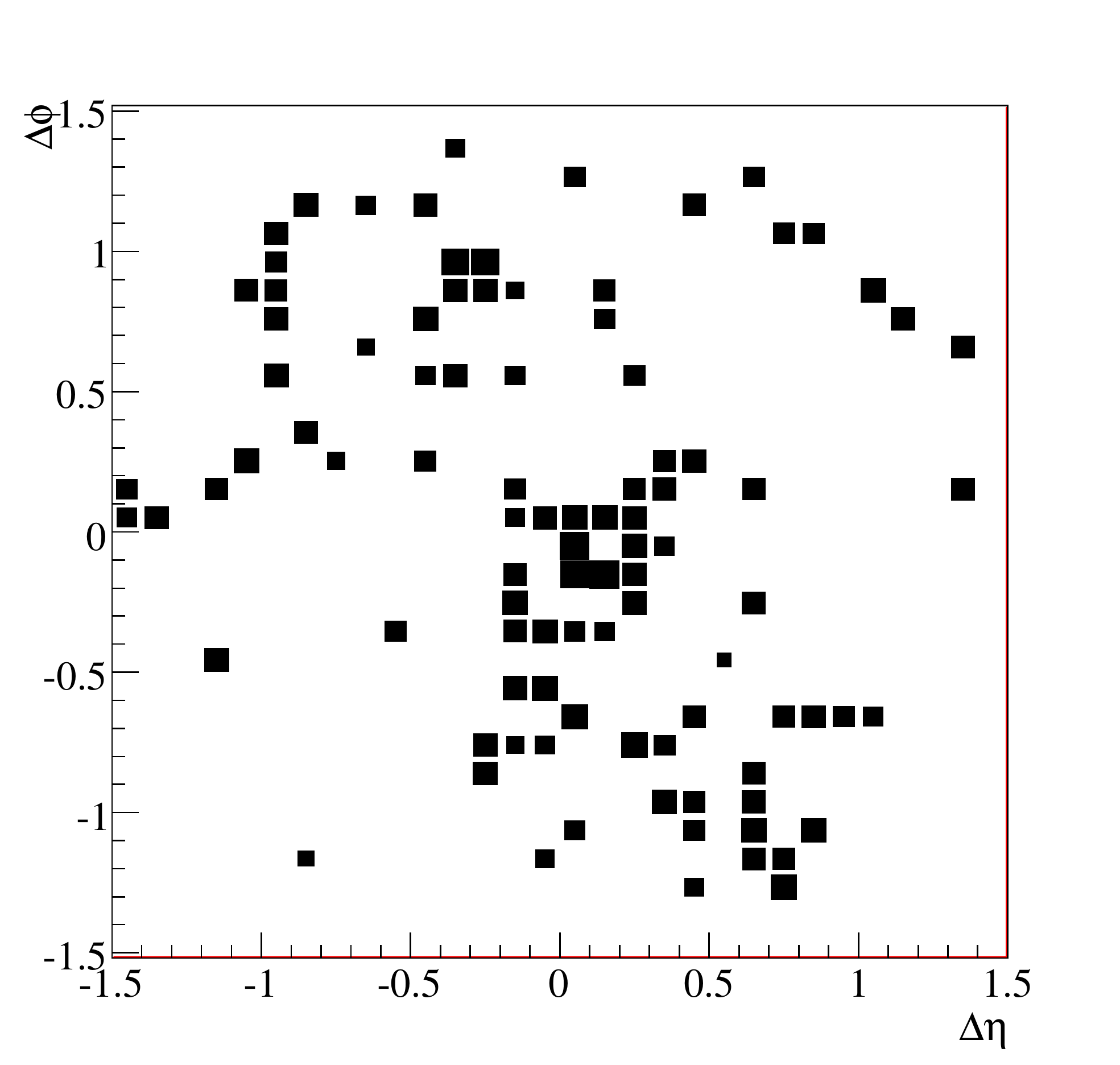}
\includegraphics[scale=0.32]{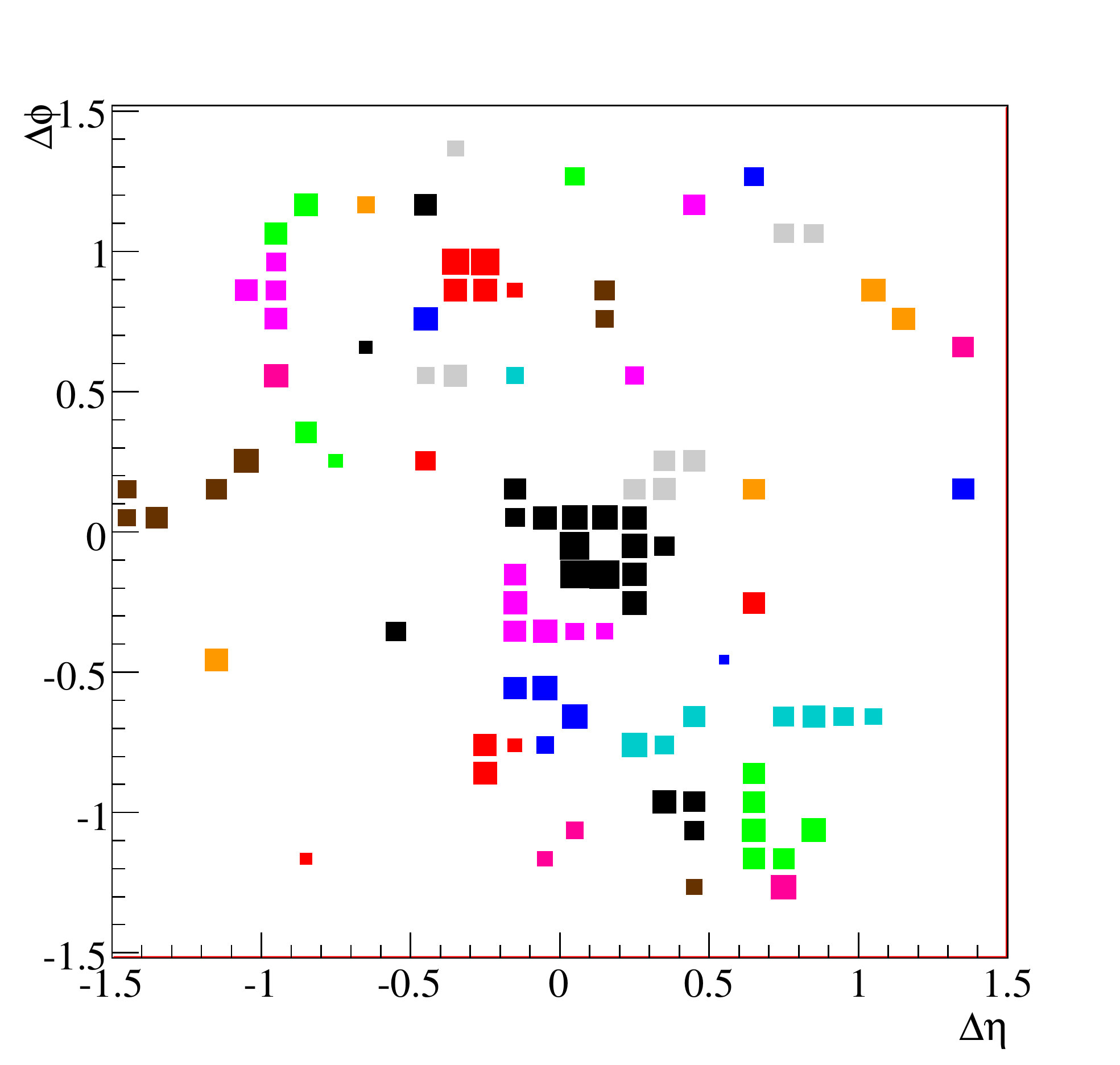}\\
\includegraphics[scale=0.32]{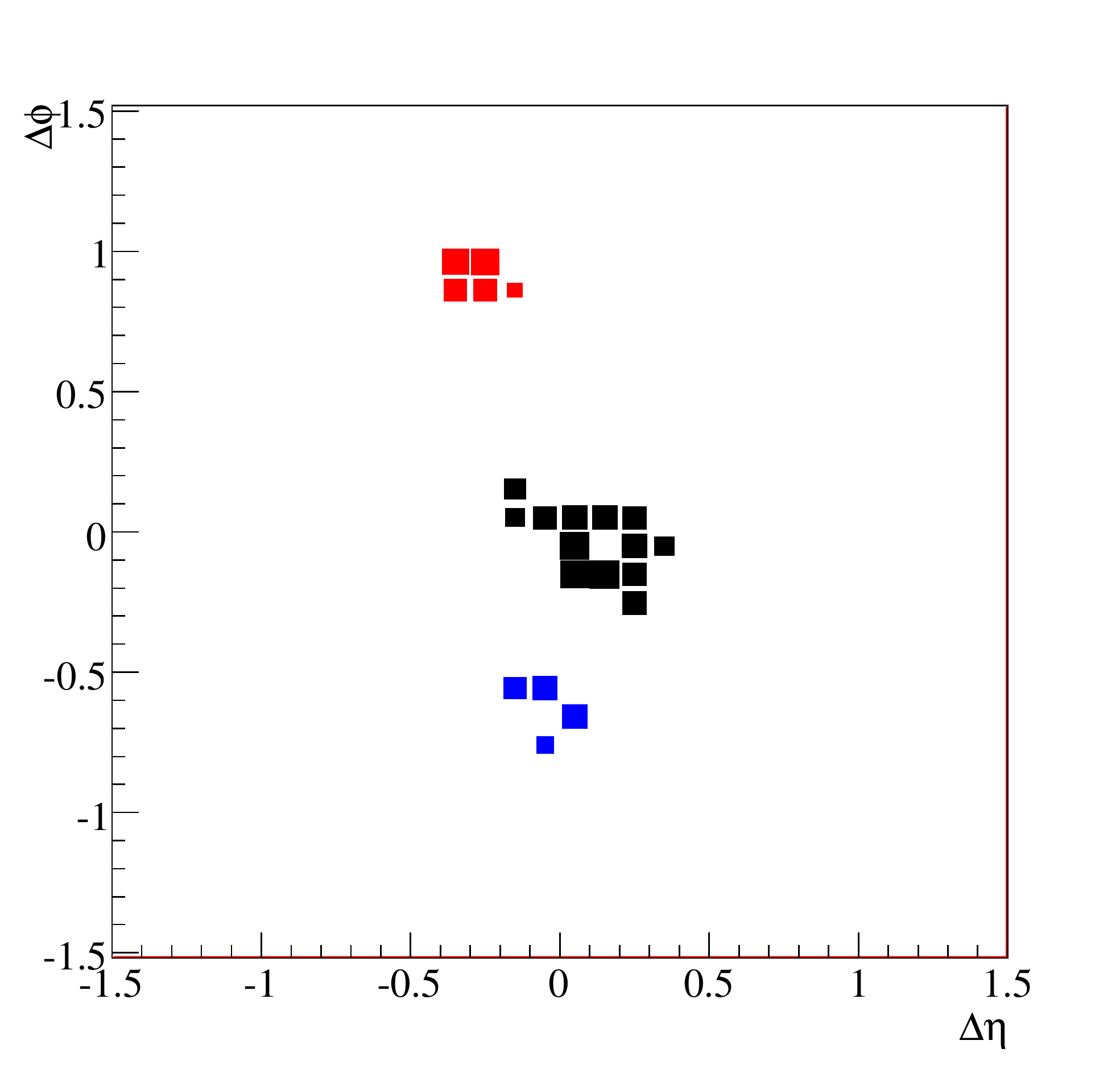}
\includegraphics[scale=0.32]{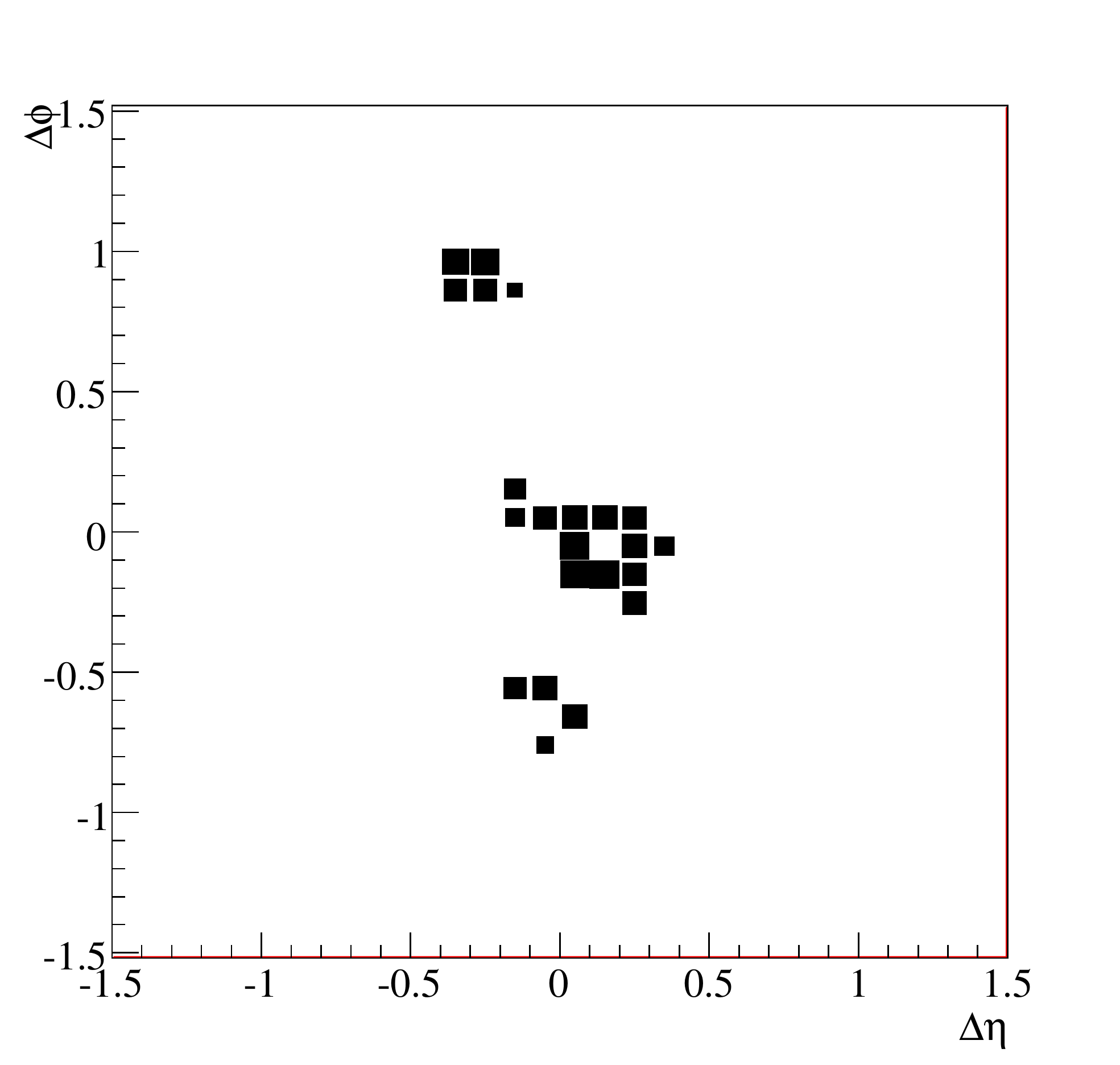}
\caption{Step by step illustration of the jet trimming procedure.  Proceeding from left to right, top to bottom, 
we show a jet as initially clustered (using anti-$k_T$ with $R_0=1.5$), 
the constituent $k_T$ subjets with $R_{\rm sub}=0.2$, the subjets surviving the $p_{Ti} < f_{\rm cut}\cdot p_T$ cut
(where $f_{\rm cut} = 0.03$), and the final trimmed jet.  To make the figure easier to read, the area of each cell is 
proportional to the log of the cell's $p_T$.}
\label{fig:trimming_step_by_step_log}
}

\FIGURE[t]{
\includegraphics[scale=0.32]{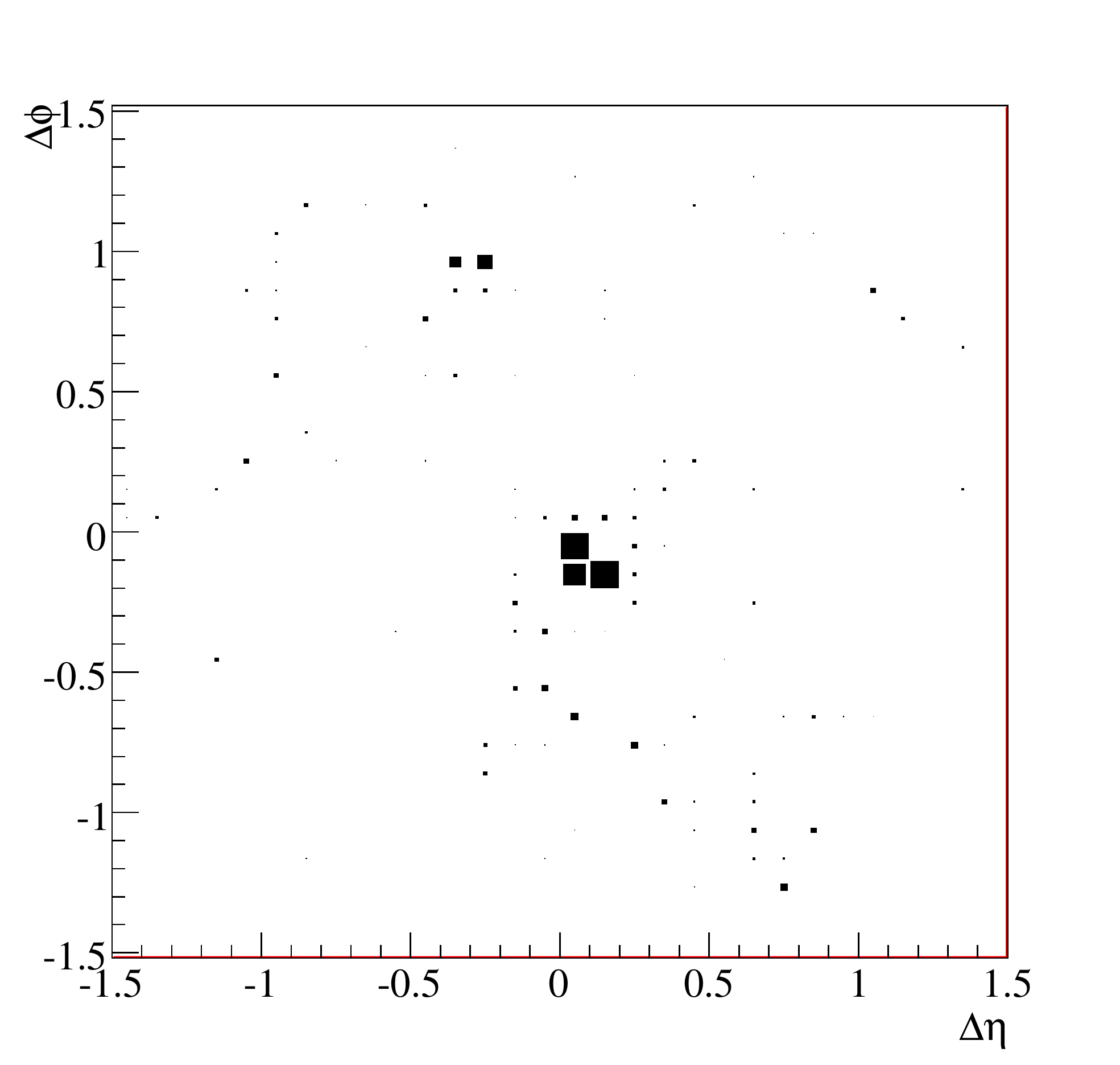}
\includegraphics[scale=0.32]{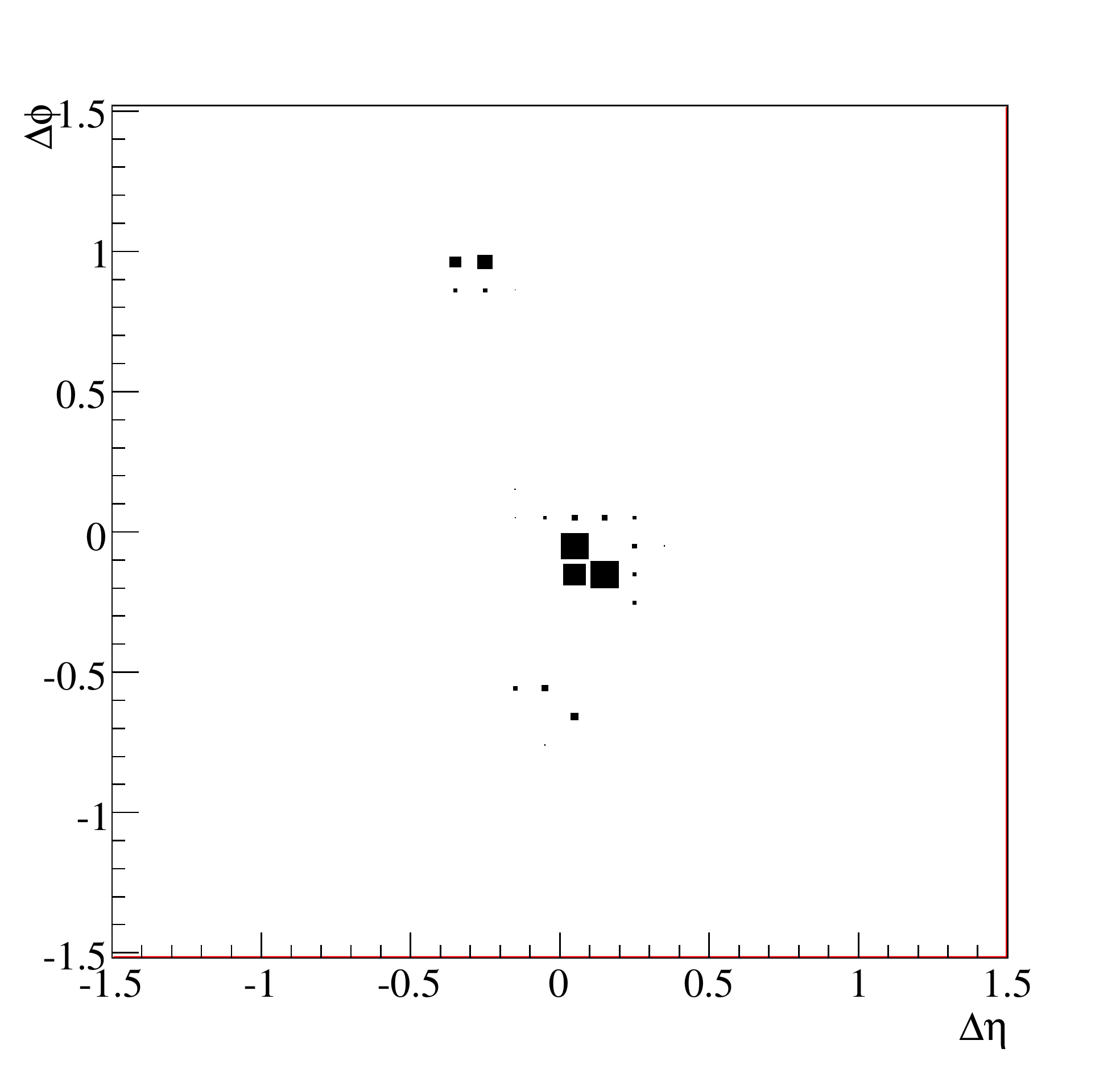}\\
\includegraphics[scale=0.32]{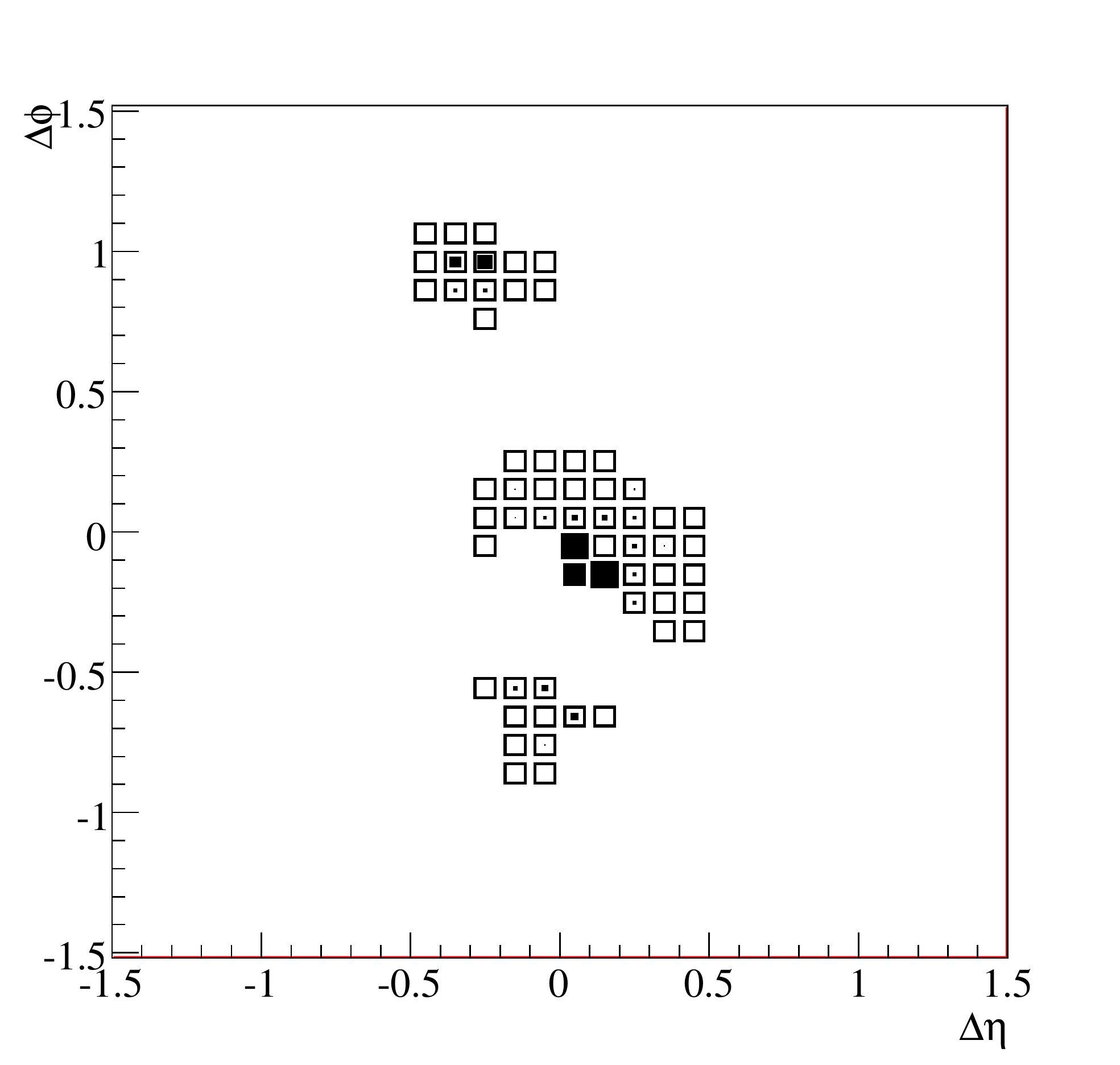}
\caption{The jet of \Fig {fig:trimming_step_by_step_log} before (top left) and after (top right) trimming using a linear scale where a cell's area is proportional to its $p_T$.
Also shown in the lower panel is the catchment area of the jet \cite{Cacciari:2008gn},  where the empty black squares indicate cells that would have been clustered in the final trimmed jet if all cells were given an infinitesimal amount of radiation.  As we will discuss more in \Sec{sec:jetarea}, the jet's area has been dramatically reduced, here to around $8\%$ of its untrimmed value.}
\label{fig:trimming_step_by_step_linear}
}

The precise jet definition used in step 1 is largely irrelevant for the jet trimming procedure.  In \Sec{sec:results}, we will 
trim two different jet algorithms, anti-$k_T$~\cite{Cacciari:2008gp} and VR~\cite{Krohn:2009zg}, finding improvements in reconstruction 
with both. 

The jet definition used in step 2, however, is more important as it determines how the subjets are found.  
We use the $k_T$ algorithm~\cite{Catani:1993hr,Ellis:1993tq} rather than a Cambridge-Aachen~\cite{Dokshitzer:1997in,Wobisch:1998wt} 
or anti-$k_T$ algorithm~\cite{Cacciari:2008gp}, because subjets formed by the $k_T$ algorithm tend to better share the energy 
between subjets.  That is, imagine that the dominant FSR depositions in a seed jet cannot be contained 
within a single subjet of size $R_{\rm sub}$.  In such circumstances the anti-$k_T$ algorithm, which clusters 
radiation from hardest to softest, will tend to create imbalanced subjets by allocating most of the energy 
to one subjet, making it more likely that the weaker subjet will be discarded when the softness criteria is 
applied.  As the $k_T$ algorithm clusters from softest to hardest, it is more likely to yield a equitable distribution 
of energy between the subjet that contain FSR, making them less likely to be discarded by the trimming procedure.

Finally, we must select a $\Lambda_{\rm hard}$ to set our criterion of hardness when judging a subjet's $p_T$.   This is a non-trivial choice, as different kinematical configurations call for different  scales, and the difference in reconstruction from different scale choices can be large.  To illustrate this, in \Sec{sec:results} we will present two possibilities for $\Lambda_{\rm hard}$: the seed 
jet's $p_T$, and the effective mass of the event (i.e.\ the scalar sum of the transverse momenta: $H=\sum p_T$).  While we have only considered the simplest two scale choices, it would be interesting to investigate more complicated methods to see if additional gains could be realized.

\subsection{Comparison to Previous Methods}

As argued before, most techniques useful in removing contamination from the jets of boosted heavy particles keep a fixed number $N_{\rm cut}$ of hard subjets.  
  To enable an apples-to-apples comparison of $f_{\rm cut}$ vs.\ $N_{\rm cut}$, we will simply take the jet trimming algorithm above, replacing step 3 with:
\begin{enumerate}
  \setcounter{enumi}{2}
  \item Sort the subjets according to $p_T$ and discard the contributions of those softer than the $N_{\rm cut}$-th hardest subjet.
\end{enumerate}
This allows us to employ a condensed notation in discussing various trimming procedures. We can denote different algorithms by
\begin{equation}
{\rm alg}(\{f,N\}, \{p_T,H\}),
\end{equation}
where ``alg'' is the algorithm used to make the seed jets (here anti-$k_T$ or VR), $\{f,N\}$ specifies whether we will trim all subjets with a $p_T$ below $f_{\rm cut}\cdot \Lambda_{\rm hard}$
or include only the $N_{\rm cut}$ hardest subjets, and the final entry (only relevant for $f_{\rm cut}$ algorithms) species whether we will use the $p_T$ of the seed jet  or the effective mass of the event to set $\Lambda_{\rm hard}$.  

In addition to this fixed $N_{\rm cut}$ algorithm---which we believe represents the most advantageous application of previous techniques for boosted objects to the study of light parton jets---we will also include a direct implementation of an algorithm from an earlier study.  We will present results using the jet filtering technique of \Ref{Butterworth:2008iy} (labelled {\it Filtering} below) which takes a jet and only includes those constituents that fit into the two hardest C/A subjets formed from cones of size $R_0/2$.  

We were also interested in testing the jet pruning procedure of \Ref{Ellis:2009su}, since it could be considered as something of a middle ground between trimming  and the cleaning methods that cut on a fixed number of subjets.\footnote{The implementation of pruning we tested was \texttt{FastPrune 0.3.0}~\cite{Vermilion:FastPrune}.}   Pruning functions by attempting to remove spurious mergings in the clustering sequence: after a seed jet is formed, its constituents are reclustered using the $k_T$ or C/A algorithm, and if the jet algorithm attempts to merge widely separated ($\Delta R > 2 R_{P} m_J/p_T$) four-momenta with a large $p_T$ hierarchy ($z<z_{\rm cut}$)\footnote{Here $z=\min[p_{T_a},p_{T_b}]/p_{T_{\rm tot}}$ for four-momenta $p_a$ and $p_b$ with $p_{{\rm tot}}=p_a+p_b$.} then the merging is rejected, and the softer of the two four-momenta deleted.  Pruning is most effective at removing spurious mergings from the later stages of clustering (i.e. right before the jet is complete), which is precisely what it should do to reconstruct a boosted heavy particle.

However, in our studies on pruning light quark jets we found at best only a marginal improvement in reconstruction. This occurred when the optimized value of $R_P$ was essentially zero, meaning that the $z_{\rm cut}$ criteria was being applied at every stage of the reconstruction.\footnote{Because the improvements we see with pruning are small, and occur at parameters at which the procedure is uncomfortably sensitive to the calorimeter segmentation, we will not report them in the next section.}    We suspect that the reason pruning is not working well in this context is that far enough down the line in the parton shower, there is no longer a clear scale separation between FSR and contamination.  So while pruning employs a relative $p_T$ cut (as in jet trimming), it appears to be most effective when employed on a jet with a fixed, small number of hard subjets.  It is an open question whether pruning techniques might be modified to successfully clean light quark jets.

\section{Results}
\label{sec:results}

We now apply the above jet trimming procedure to two examples in different kinematic regimes:  heavy resonance reconstruction, and a two-step decay chain.   Our goal is justify the use of trimming, show that  it is advantageous to use a trimming procedure specifically designed for jets from light partons, and to see how different measures of $\Lambda_{\rm hard}$ can change the reconstruction of the trimmed event.  Unlike \Sec{sec:benefits}, here we will include both the effects of ISR/MI and event pileup.

In both examples, we will find that employing any sort of trimming procedure leads to an improvement in reconstruction.   However, in going from an algorithm designed for boosted heavy particles to one specifically aimed at light parton jets, we can realize significant additional gains.  Further, using a measure of hardness well suited to the kinematics of an event can make almost as big a difference in reconstruction as to the decision to trim in the first place.  
 
Our results confirm our intuitions from \Sec{sec:benefits} that trimming partially resolves the jet-size/contamination tradeoff.  For the anti-$k_T$ algorithms, the optimal $R_0$ value in the trimmed sample is systematically larger than the optimal $R_0$ value in the untrimmed sample. Similar conclusions hold for VR, with the jet size parameter $\rho$ being larger in the trimmed samples.\footnote{In VR algorithms, the radius of a jet is approximately $R\approx \rho/p_T$, where $p_T$ is the jet's transverse momentum.}  We will find that background dijet distributions are not increased through the use of a large initial radius, and may even be reduced in some cases.  Finally, as expected, the active jet area~\cite{Cacciari:2008gn} is substantially smaller in the trimmed sample.

To quantify reconstruction performance, we will fit reconstructed invariant mass distributions to a sum of two distributions (similar to what was done in Ref.~\cite{Ellis:2009su}):
\begin{align}
\label{eq:sig_dist}
&S(m)=\alpha\left[\frac{1+\beta(m-M)}{(m^2-M^2)^2+\Gamma^2 M^2}\right] ,\\
\label{eq:bg_dist}
&B(m)=\delta+\gamma/m , 
\end{align}
where $\delta$ and $\gamma$ are restricted to be $\geq 0$.  Here $S(m)$ is a skewed Breit-Wigner distribution and $B(m)$ is a background-like falling distribution.  We quantify signal reconstruction via the measure
\be
\label{eq:recon_measure}
\Delta \equiv S(M)= \frac{\alpha}{\Gamma^2 M^2},
\ee 
i.e.\ the peak height of the $S(m)$ curve.  While other measures of reconstruction performance would be equally reasonable, this measure favors algorithms reconstructing a tall $S(m)$ of narrow width, and has the advantage of not introducing any arbitrary parameters beyond the fitted functional form.  Note that this reconstruction measure does not attempt to reward algorithms that get the right peak position, and we will see a corresponding systematic invariant mass shift in using trimmed jets.  

For simplicity of discussion, we only consider processes with initial/final state gluons.   From \Tab{tab:inPrincipleBest}, we see
that improvements are certainly possible when these are replaced with light quarks, and all of our conclusions regarding the optimal trimming method will hold there as well.  It is important to remember, though, that quarks have a lower effective color charge then gluons and thus produce less QCD radiation. Thus, for light quarks one expects (and we found) a diminished optimal untrimmed jet radius and a lower
potential improvement achievable through trimming.

Finally, one should keep in mind that while the improvements we find are the result of well understood physical effects, the precise values of the trimming parameters will change somewhat when the Monte Carlo tuning is adjusted to account for LHC data.  Thus, while the parameters below will provide a reasonable guide to what should be used at the LHC, the exact values will need to be inferred from a iterative process of Monte Carlo tuning to standard candles.

\subsection{Heavy Resonance Decays}
The simplest test of a jet algorithm is how it reconstructs a heavy resonance decaying to the two jets.  As in \Sec{sec:benefits}, we use the process $gg\rightarrow \phi\rightarrow gg$ where $\phi$ is a color octet scalar with $m_\phi = 500~{\rm GeV}$.

\TABLE[t]{
\parbox{\textwidth}{

\begin{center}
\begin{tabular}{|c|c|c|c|c|c|c|}
\hline
\bf	& Improvement&$f_{\rm cut}$, $N_{\rm cut}$& $R_{\rm sub}$ &$R_0$,  $\rho$ &$\Gamma$ [GeV] & $M$ [GeV]\\
\hline
anti-$k_T$ & - & - & - & $1.0^\ast$ & $71$ & $522$\\
anti-$k_T$ ($N$) & $40\%$ & $5^\ast$  & $0.2^\ast$ & $1.5^\ast$ & $62$ & $499$\\
anti-$k_T$ ($f$, $p_T$) &  $59\%$ &$3\times 10^{-2\ast}$ &0.2 & 1.5 &  $52$ & $475$ \\
anti-$k_T$ ($f$, $H$) &  $61\%$ &$1\times 10^{-2\ast}$ &0.2 & 1.5 &  $50$ & $478$ \\
\hline
VR & $30\%$ & - & - &$200^\ast~{\rm GeV}$& $62$ & $511$\\
VR ($N$)	&$53\%$&$5$&0.2&$275^\ast~{\rm GeV}$&	$53$&	$498$\\
VR ($f$, $p_T$) & $68\%$ & $3\times 10^{-2}$ & $0.2$ & $300^\ast~{\rm GeV}$&  $49$ & $475$\\
VR ($f$, $H$) & $73\%$ & $1\times 10^{-2}$ & $0.2$ & $300^\ast~{\rm GeV}$&  $47$ & $478$\\
\hline
Filtering& $27\%$ & 2  & $R_0/2$ & $1.3^\ast$ & $61$ &$515$\\
\hline
\end{tabular}
\end{center}
\caption{
\label{tab:trimming_dijet}
Comparison of dijet resonance reconstruction using trimmed and untrimmed algorithms.  
The first column specifies the algorithm, the second lists the change in $\Delta$  over untrimmed anti-$k_T$ (second row), the third lists the relevant trimming parameters, the fourth contains 
the subjet radius, the fifth the seed jet parameters, the sixth the fitted width, and the seventh the fitted mass.  
For each algorithm, we have optimized those parameters denoted by a $^\ast$, while the rest have remained fixed.}
}}

The results of this reconstruction are presented in  \Tab {tab:trimming_dijet}.  Here we are interested primarily in two different comparisons: untrimmed algorithms versus those trimmed
using an $f_{\rm cut}$ (so as to measure the full potential for improvement in reconstruction), and those trimmed using an $N_{\rm cut}$ to those using an $f_{\rm cut}$.  Now, the more parameter choices
one optimizes in an algorithm the more that algorithm stands to gain from arbitrary statistical fluctuations.  To guard against this and ensure that the first comparison 
above is fair, we fully optimize the anti-$k_T$~($N$) algorithm, using the resulting best choices of $R_{\rm sub}$ and $R_0$ as inputs to our optimization of 
anti-$k_T$~($f$), for which we only optimize a single parameter: $f_{\rm cut}$.  The result is a fair comparison of untrimmed algorithms to those trimmed with an $f_{\rm cut}$, and 
a comparison of $N_{\rm cut}$ to  $f_{\rm cut}$ trimming where $N_{\rm cut}$ trimming is given a statistical advantage.\footnote{For the VR algorithms we will take the anti-$k_T$ optimized $R_0$, $f_{\rm cut}$, and $N_{\rm cut}$ as 
inputs ($R_0$ will set $R_{\rm max}$) and optimize the $\rho$ parameter.} 

Several algorithms and trimming procedures are presented in \Tab {tab:trimming_dijet}.  We have included untrimmed anti-$k_T$, anti-$k_T$ with a cut on the momenta of  $k_T$ subjets (set relative to both the jet's $p_T$ and the event's effective mass), anti-$k_T$ with a fixed number of $k_T$ subjets, and for comparison with previous techniques anti-$k_T$ with two C/A subjets of half the seed jet radius (i.e.\ the filtering procedure of \Ref{Butterworth:2008iy}).  Both trimmed and untrimmed VR jets are also included.   In \Fig{fig:reconstrestrimcom}, we display the reconstructed $\phi$ mass using both trimmed and untrimmed anti-$k_T$ and VR algorithms.

\FIGURE[t]{
\includegraphics[scale=0.32]{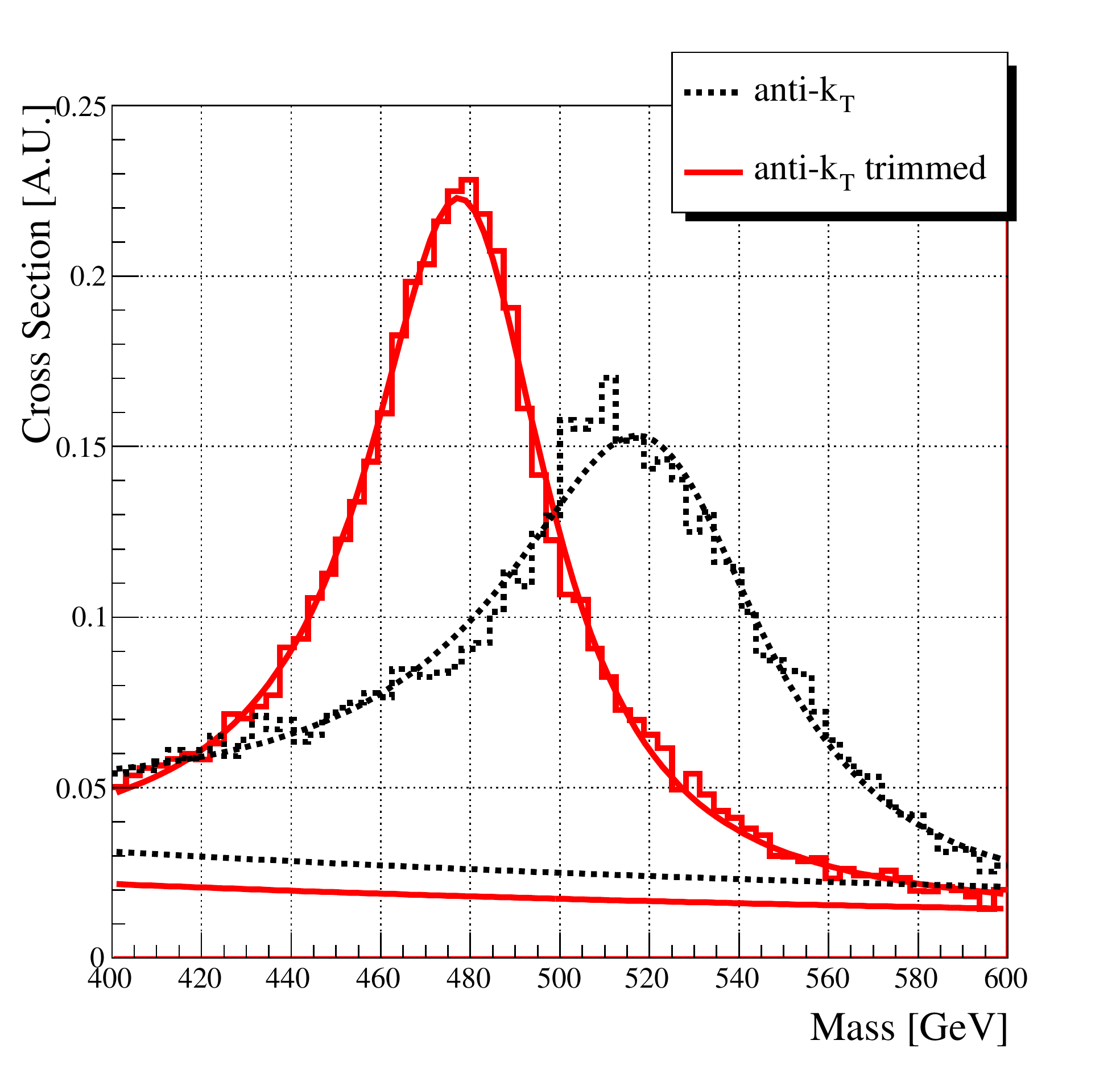}
\includegraphics[scale=0.32]{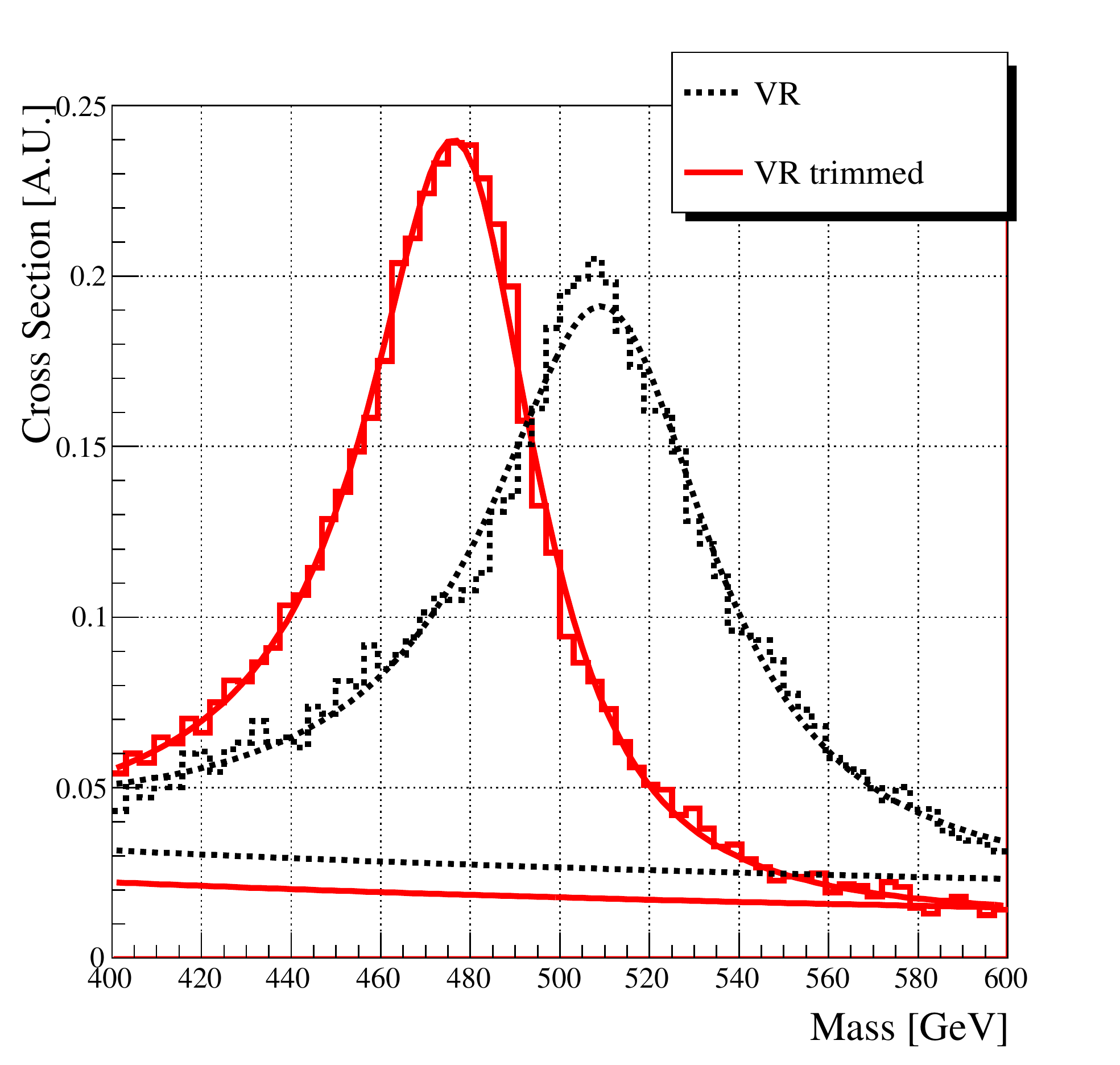}
\caption{Dijet resonance reconstruction with and without trimming using the anti-$k_T$/VR and anti-$k_T$/VR ($f$, $H$) algorithms.   The algorithm parameters are those that optimize the $\Delta$ measure of \Eq {eq:recon_measure}, as listed in  \Tab {tab:trimming_dijet}.  The upper curves are fitted to the sum of $S(m)$ and $B(m)$ from Eqs.~(\ref{eq:sig_dist}) and~(\ref{eq:bg_dist}), while the lower curves display the contribution of $B(m)$.}
\label{fig:reconstrestrimcom}
}

We see that trimming 
of any sort is useful in reconstruction.  However, the difference between trimming techniques is apparent.
By using an algorithm with a $p_T$ cut determined as a fraction of the original $p_T$ (i.e.\ the samples
whose trimming is parameterized by an $f_{\rm cut}$) we are able to see significant gains beyond what 
is possible using a fixed number of subjets.  This reflects the fact that the structure of the jet from a light parton
is not known a-priori, unlike the jets from boosted heavy particles, so it is advantageous to trim with a direct subjet $p_T$ cut.
We further note that at this stage, the difference between using $H$ and $p_T$ to set $\Lambda_{\rm hard}$ makes only a small difference in reconstruction, reflecting the fact that for dijet events $p_T\approx H/2$.  Below, we will see that the situation will change in more complicated event topologies.

Before continuing, we remark that in \Fig{fig:reconstrestrimcom}, the dijet invariant mass distribution is systematically shifted to lower values through the effects of jet trimming.  
This is to be expected, given that the trimming procedure will necessarily result in some accidental removal of FSR.  To understand the size of the effect, note that in \Tab{tab:trimming_dijet} we find an optimized $f_{\rm cut}$ of around $3\%$ when we cut on the  subjet's $p_T$ relative to that of the seed jet, and that the optimal $N_{\rm cut}$ for fixed-number cleaning is 5. Since the pattern of QCD radiation from a light parton 
ensures us that the subjets follow a strong $p_T$ hierarchy, we should only expect one or two subjets to be slightly below the $3\%$ $p_T$ cut we have imposed.  This is enough to account for the roughly $5\%$ shift in $M$ that we observe.

\subsection{Longer Decay Chains}
Next, we consider the production channel $gg\rightarrow X\rightarrow YY\rightarrow gggg$ where $m_X = 1~{\rm TeV}$ and $m_Y = 300~{\rm GeV}$.  This sample is qualitatively  different from the dijet reconstruction in two ways: the final state is more crowded, and the final 
state jets can vary widely in $p_T$ within the same event.  The results from this reconstruction are presented in \Tab {tab:trimming_quartjet}, and the resulting $m_X$ and $m_Y$ distributions are
plotted in \Fig{fig:reconstrestrimcom_quart} and \Fig {fig:reconstedy}, respectively.

\TABLE[t]{
\parbox{\textwidth}{
\begin{center}
\begin{tabular}{|c|c|c|c|c|c|c|}
\hline
\bf	& Improvement&$f_{\rm cut}$, $N_{\rm cut}$& $R_{\rm sub}$ &$R_0$,  $\rho$ &$\Gamma$ [GeV] & $M$ [GeV]\\
\hline
anti-$k_T$ & - & - & - & $0.8^\ast$ & $158$ &	$994$\\
anti-$k_T$ ($N$) & $12\%$ & $5^\ast$  & $0.2^\ast$ & $1.0^\ast$ & $115$ & $969$\\
anti-$k_T$ ($f$, $p_T$) &  $10\%$ &$3\times 10^{-2\ast}$ &0.2 & 1.0 &  $108$ & $941$ \\
anti-$k_T$ ($f$, $H$) &  $19\%$ &$5\times 10^{-3\ast}$ &0.2 & 1.0 &  $100$ & $944$ \\
\hline
VR & $10\%$ & - & - &$150^\ast~{\rm GeV}$& $157$ & $979$\\
VR ($N$) & $17\%$ & $5$ & $0.2$ & $275^\ast~{\rm GeV}$&  $115$ & $965$\\
VR ($f$, $p_T$) & $16\%$ & $3\times 10^{-2}$ & $0.2$ & $225^\ast~{\rm GeV}$&  $112$ & $938$\\
VR ($f$, $H$) & $22\%$ & $5\times 10^{-3}$ & $0.2$ & $300^\ast~{\rm GeV}$&  $101$ & $942$\\
\hline
Filtering & $6\%$ & 2  & $R_0/2$ & $0.9^\ast$ & $128$ &$969$\\
\hline
\end{tabular}
\end{center}
\caption{Comparison of the $2\rightarrow 4$ resonance reconstruction using trimmed and untrimmed algorithms.  
Reconstruction is performed by taking the invariant mass of the hardest four jets.
The first column specifies the algorithm, the second lists the change in $\Delta$  over untrimmed anti-$k_T$ (second row), the third lists the trimming parameters, the fourth contains 
the subjet radius, the fifth the seed jet parameters, the sixth the fitted width, and the seventh the fitted mass.  
For each algorithm we have 
optimized those parameters denoted by a $^\ast$, while the rest have remained fixed.}
\label{tab:trimming_quartjet}
}}

\FIGURE[t]{
\includegraphics[scale=0.32]{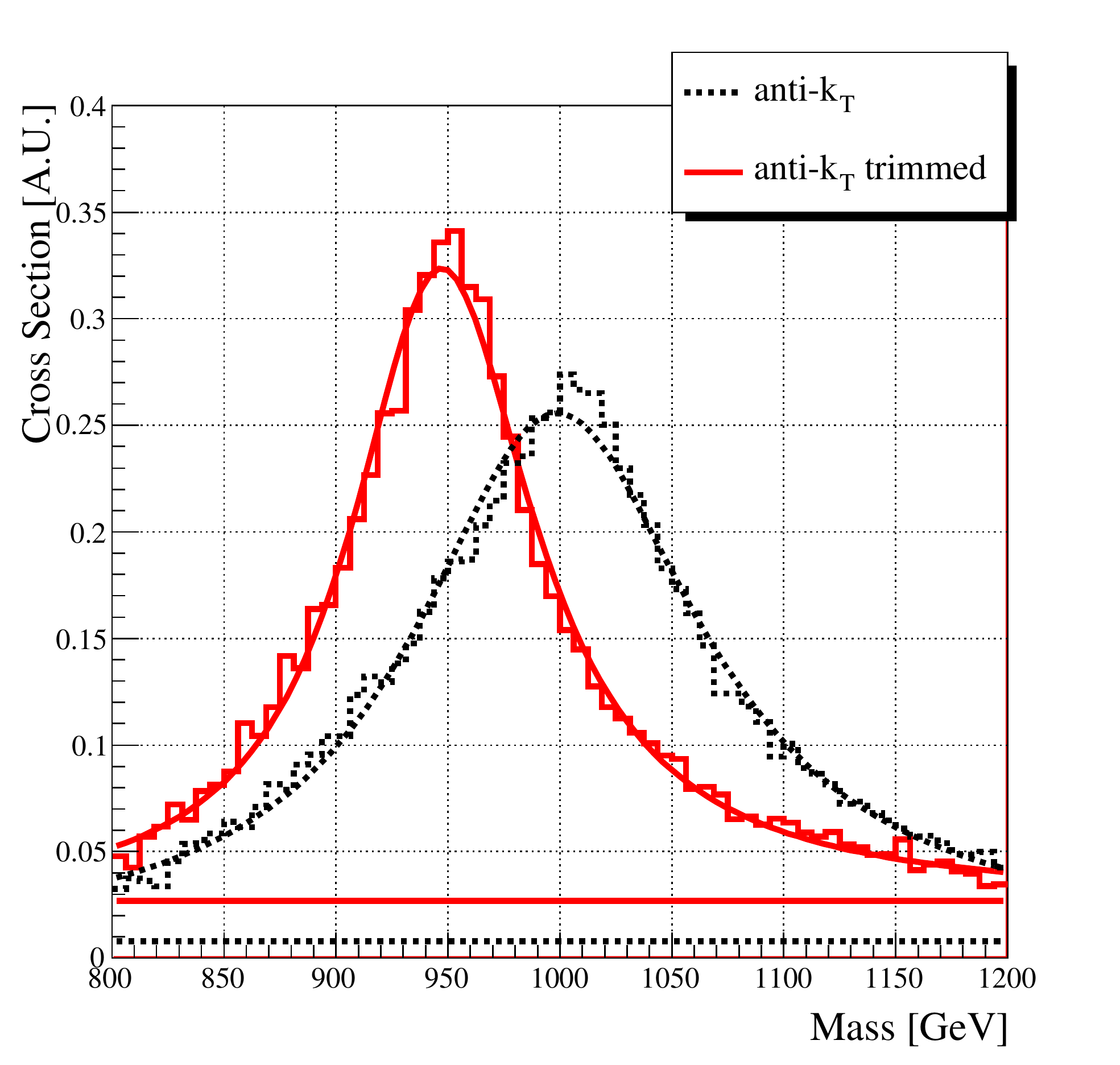}
\includegraphics[scale=0.32]{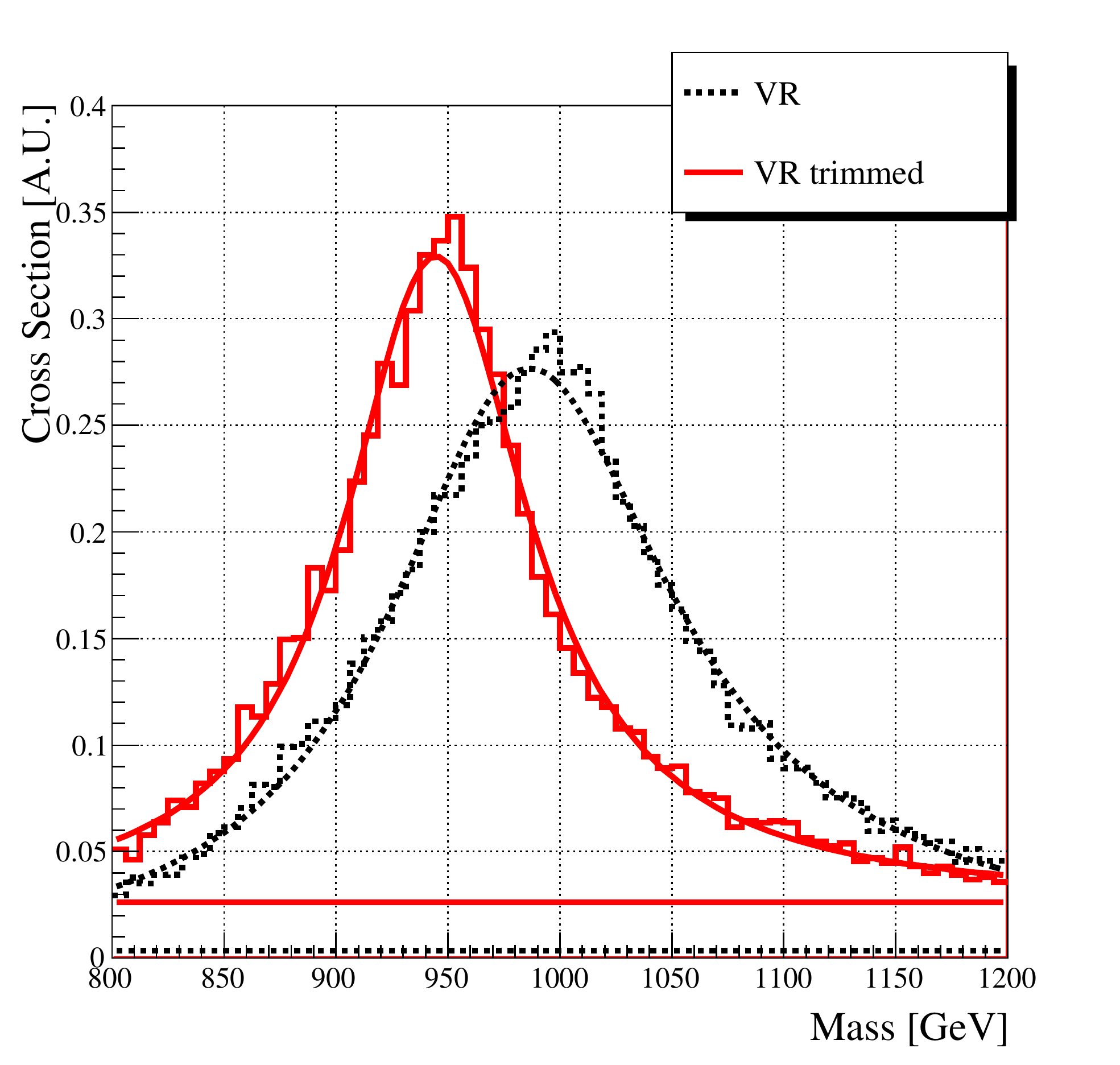}
\caption{Reconstruction of the resonance $m_X=1~{\rm TeV}$ in $gg\rightarrow X\rightarrow YY\rightarrow gggg$ with and without trimming using the anti-$k_T$/VR and anti-$k_T$/VR ($f$, $H$) algorithms.   The algorithm parameters are those optimized to those that optimize the $\Delta$ measure of \Eq {eq:recon_measure}, as listed in  \Tab{tab:trimming_quartjet}. The upper curves are fitted to the sum of $S(m)$ and $B(m)$ from Eqns.~(\ref{eq:sig_dist}) and~(\ref{eq:bg_dist}), while the lower curves represent the contribution of $B(m)$.}
\label{fig:reconstrestrimcom_quart}
}

\FIGURE[t]{
\includegraphics[scale=0.32]{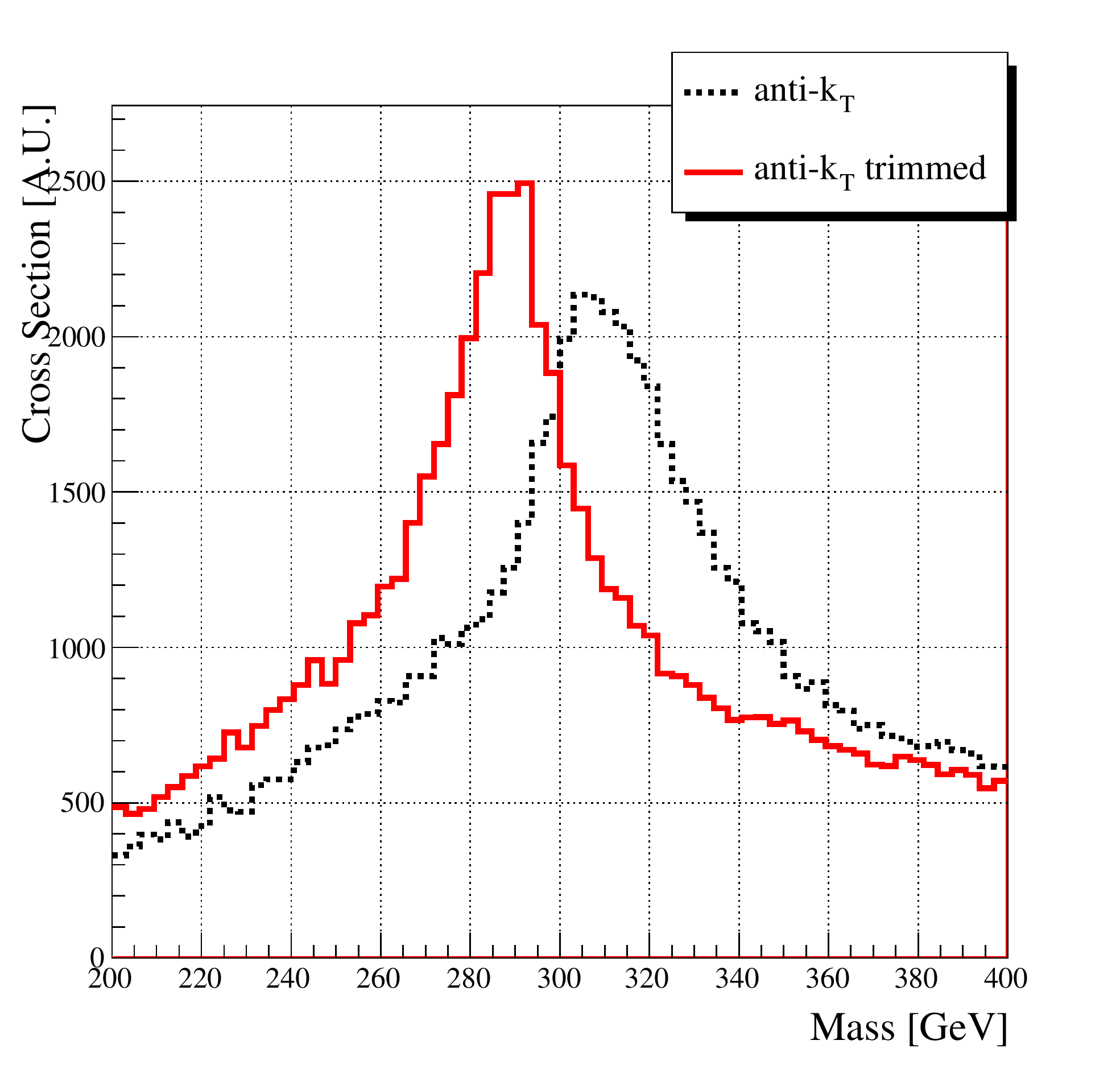}
\includegraphics[scale=0.32]{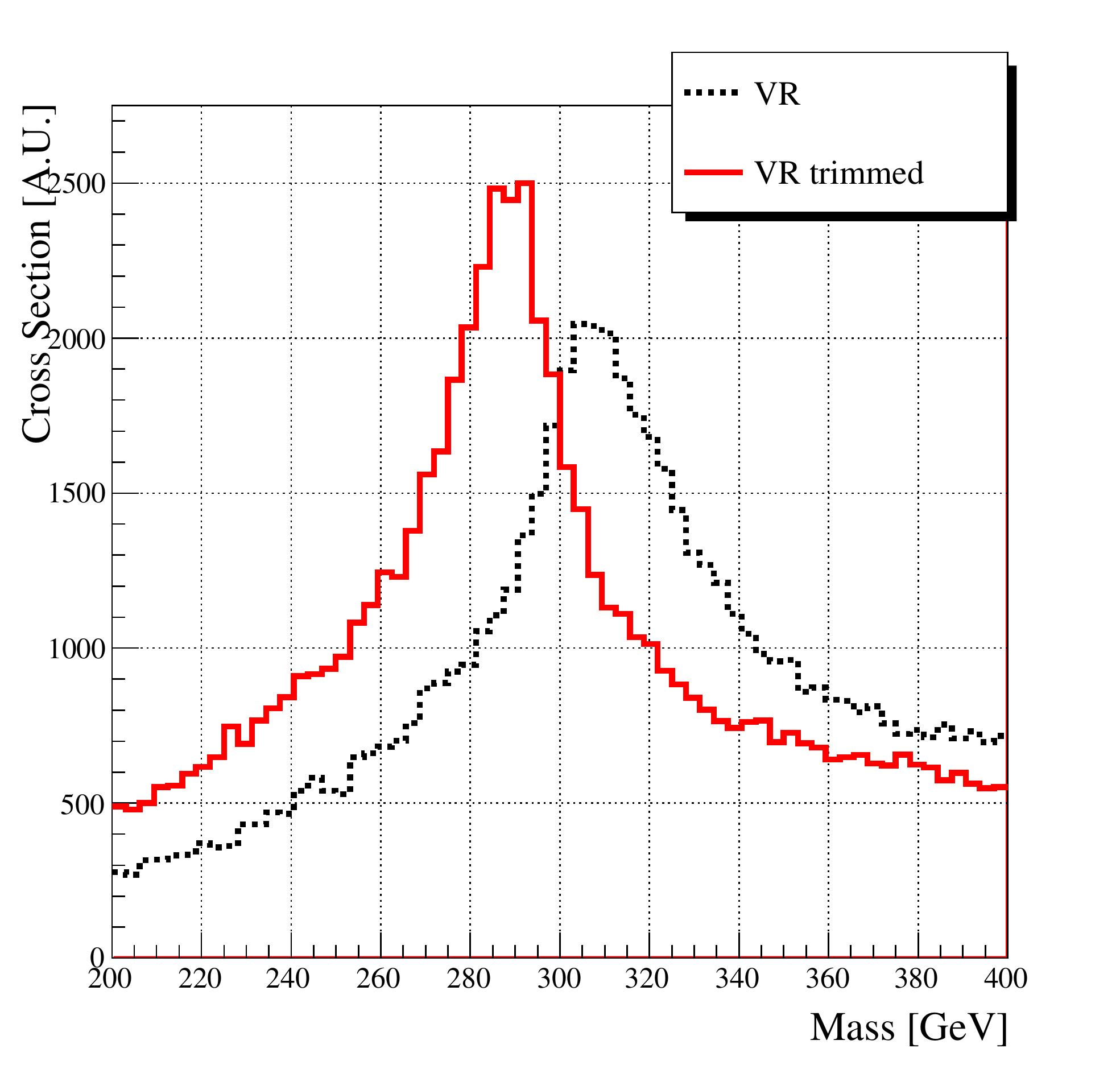}
\caption{Reconstruction of the intermediate resonance mass $m_Y=300~{\rm GeV}$ in the process $gg\rightarrow X\rightarrow YY\rightarrow gggg$ using  the anti-$k_T$/VR and anti-$k_T$/VR ($f$,$H$)  parameters of  \Tab {tab:trimming_quartjet}.  These distributions are formed by taking the four hardest jets in each event, considering the two masses from every possible $2\times 2$ partition of these jets, 
and plotting the masses from the most equitable partition (defined as the one for which $m_{\rm min}/m_{\rm max}$ is closest to one).}
\label{fig:reconstedy}
}

That the final state is crowded somewhat limits the improvements achievable from trimming.  We saw 
before in \Tab{tab:trimming_dijet} that trimming seemed to work well when the seed jets were allowed 
to grow much larger than the optimized untrimmed jets.  Here, the untrimmed jets are optimized at $R_0 = 0.8$, 
so the trimmed jets cannot grow much larger without merging with each other and ruining the reconstruction.
Despite this limitation, however, we see that valuable improvements are still possible.

More importantly, now we see that the choice of $\Lambda_{\rm hard}$ can make a significant difference in reconstruction.
When $\Lambda_{\rm hard}$ is chosen to be the effective mass of the event, reconstruction is improved beyond the case where $\Lambda_{\rm hard}$ is the seed jet $p_T$ (the improvement roughly doubles).  This is because when we let the seed jet $p_T$ determine the hard scale for each jet while using a fixed $f_{\rm cut}$, the softer jets will see little trimming (because the  minimum subjet $p_T$ is soft), while for the same reason the harder jets will see too much trimming.  The resolution, it  seems, is to simply use a global $p_T$ cutoff for each event when the signal jets are of different characteristic $p_T$ scales.

\subsection{Dijet Backgrounds}

The improvements in signal reconstruction seen so far would be of little use if jet trimming significantly increased the background as well.  After all, to see improvements in signal reconstruction we must let our seed jets cluster with a large radius, and it is possible that this could result in an unintended rise in the background distributions.

\FIGURE[t]{
\includegraphics[scale=0.32]{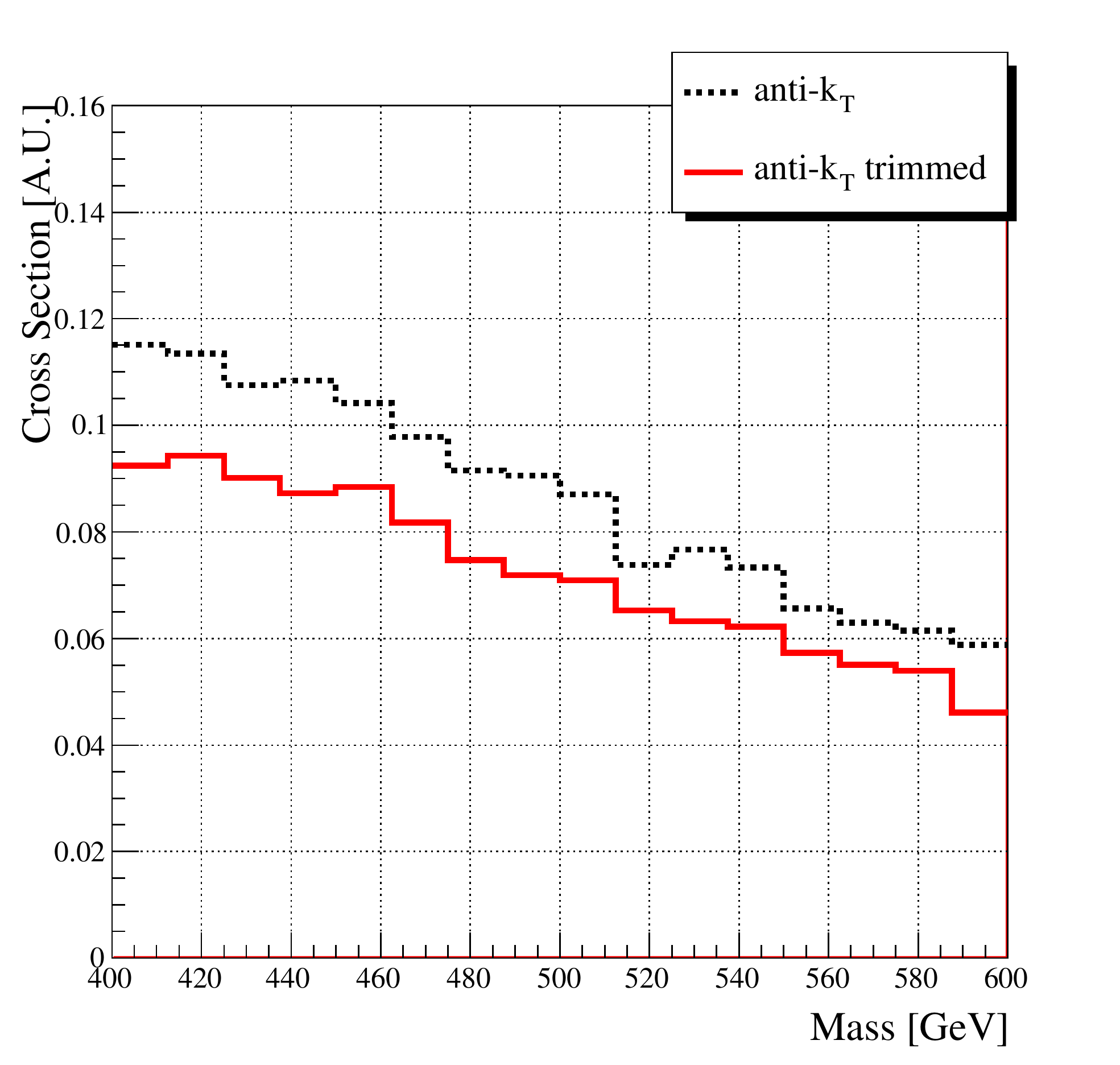}
\includegraphics[scale=0.32]{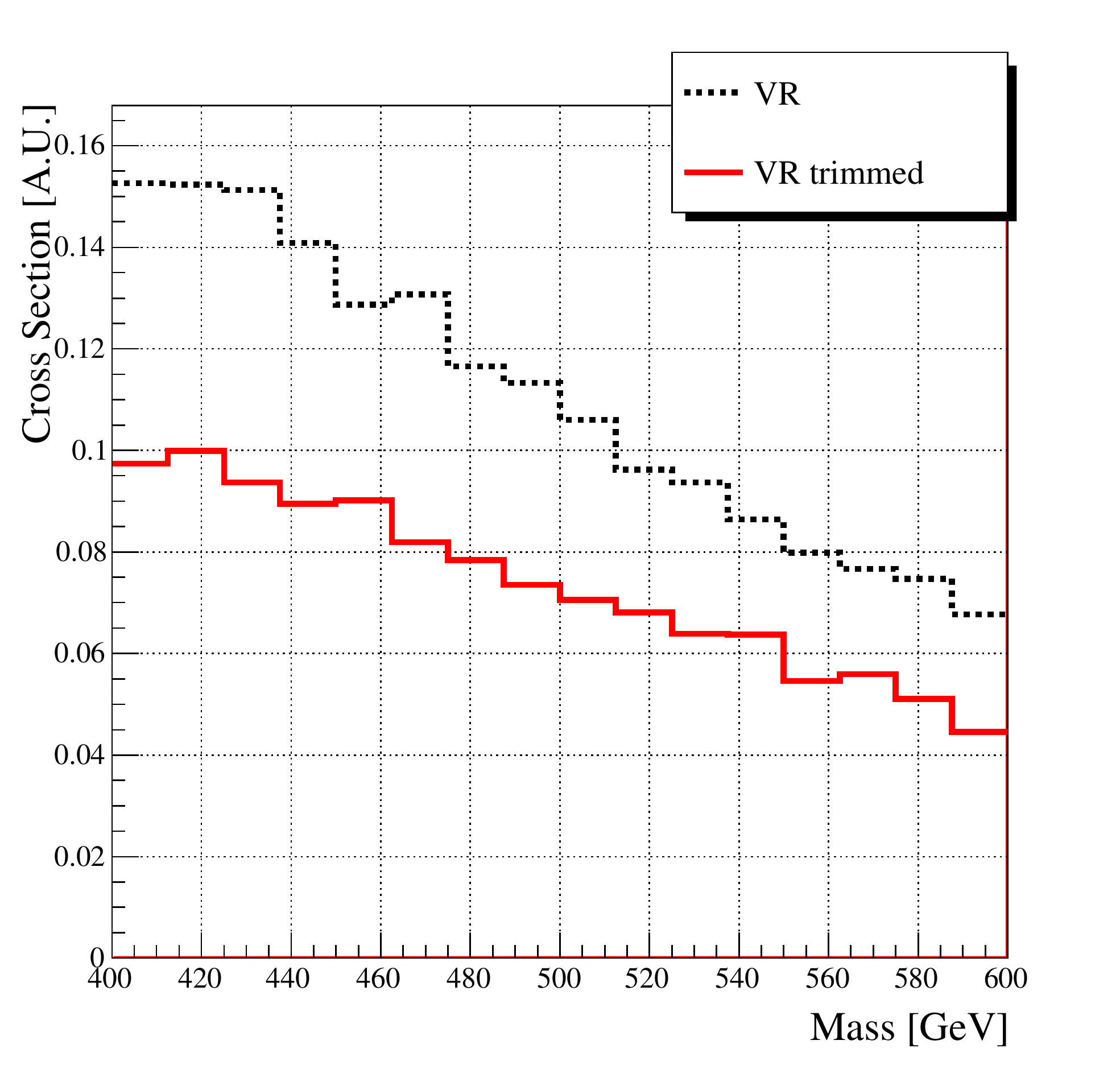}
\caption{Standard model QCD dijet background reconstruction with and without trimming using the anti-$k_T$/VR and anti-$k_T$/VR ($f$, $H$) algorithms and the optimized signal parameters from \Tab{tab:trimming_dijet}.}
\label{fig:qcddijet}
}


Fortunately, this does not seem to be the case.  In \Fig{fig:qcddijet} we present the background QCD dijet invariant mass distributions clustered using the parameters of \Tab {tab:trimming_dijet} optimized for signal reconstruction.  If anything, we see that the trimmed distributions are shifted to lower invariant mass values than the untrimmed distributions.   This is especially useful in the case of the VR jet algorithm, which on its own can distort background distributions to higher values.\footnote{In \Ref{Krohn:2009zg} we had to impose a jet-quality cut to prevent this distortion.  With jet trimming, it seems that such a cut is unnecessary.}  It is tempting to argue from this that trimming can also be useful in reducing the background, but one should be careful drawing such a conclusion as the signal position also shifts.   The precise signal and background interplay, while intriguing, is therefore likely to be highly process dependent, and requires a dedicated study.

\subsection{Jet Area}
\label{sec:jetarea}

\FIGURE[t]{
\includegraphics[scale=0.32]{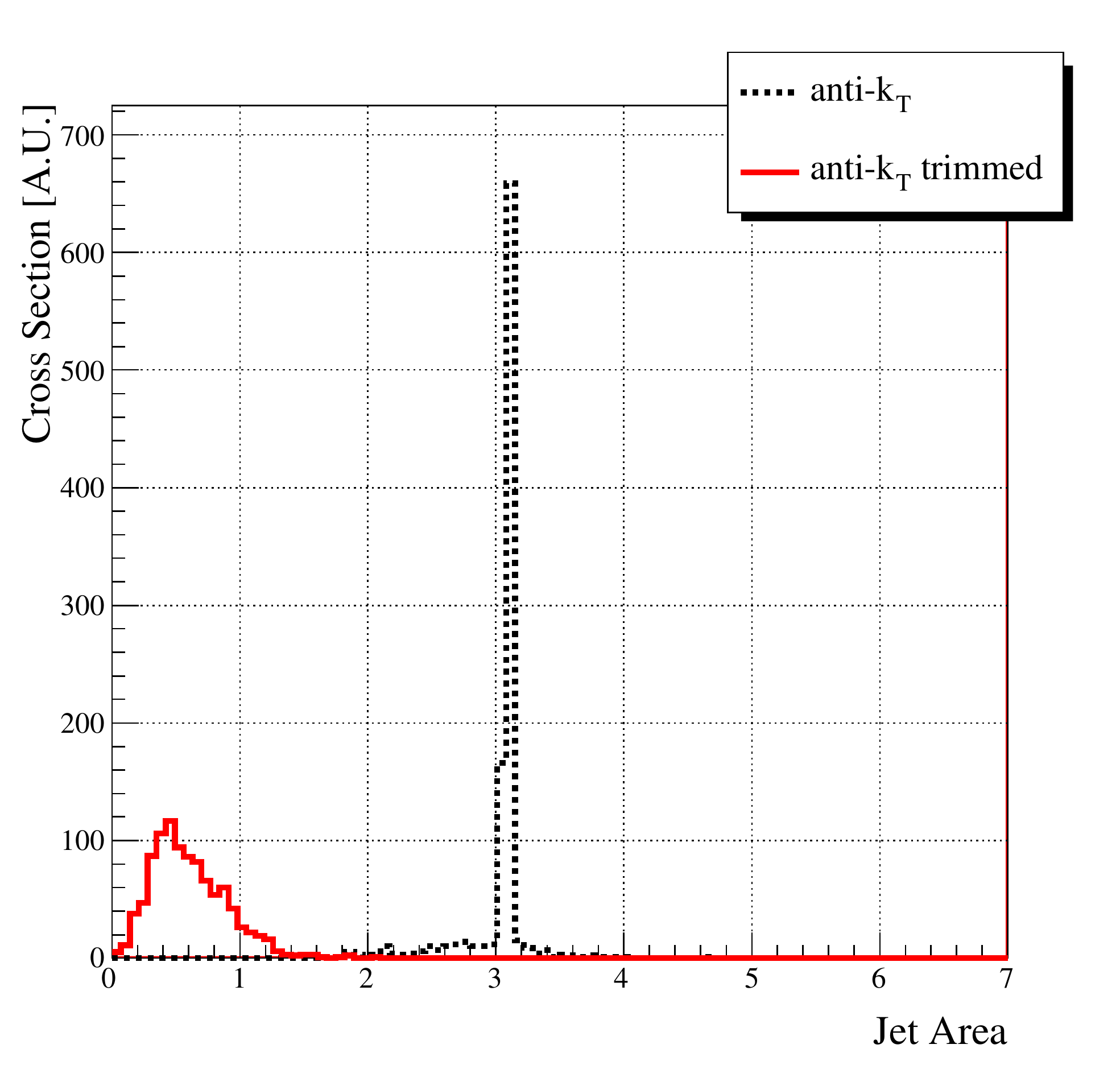}
\includegraphics[scale=0.32]{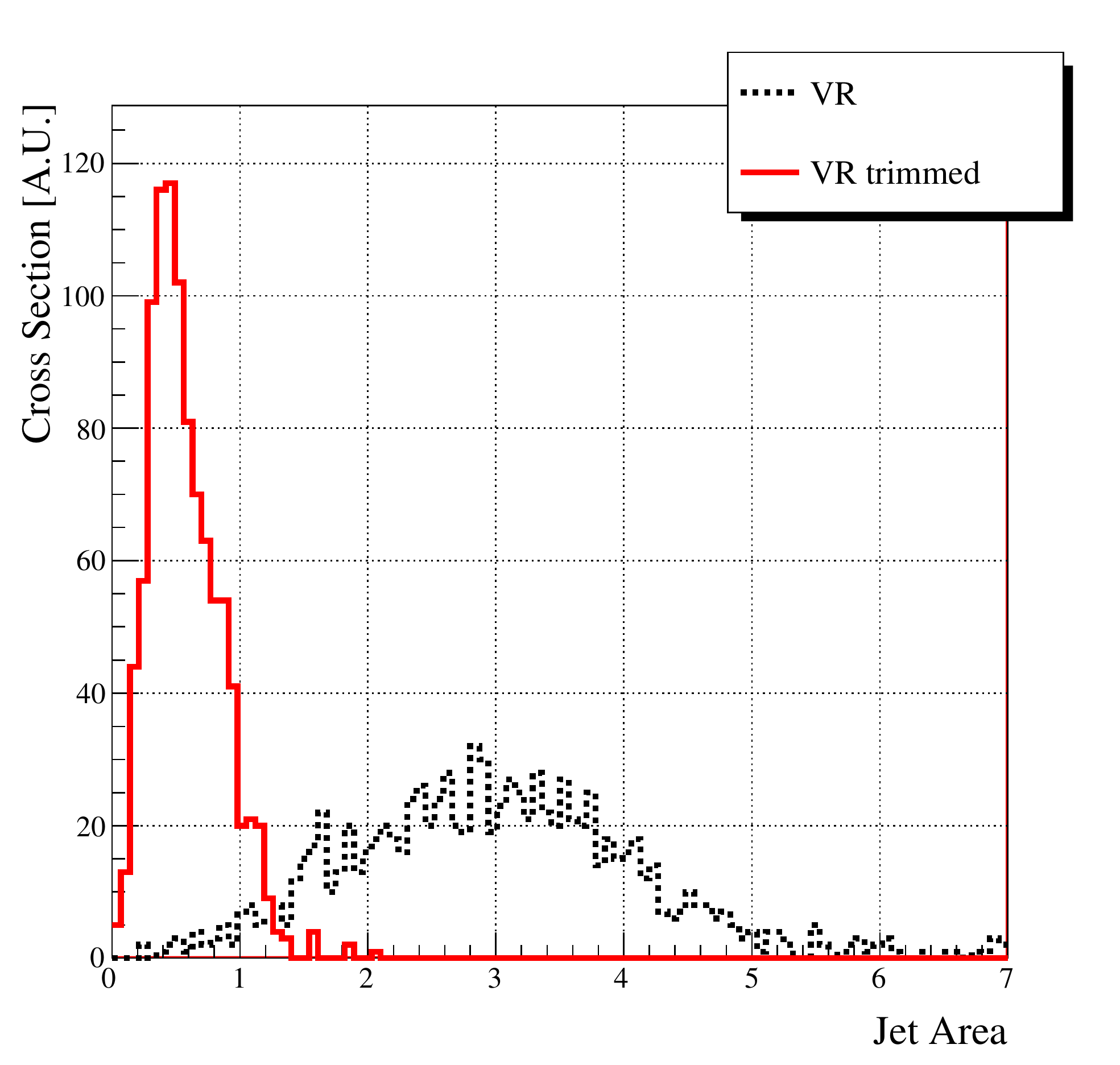}
\caption{Jet area, defined as the area of calorimeter cells clustered into a jet if each cell contains 
at least an infinitesimal about of radiation, for anti-$k_T$ vs.\ anti-$k_T$($f$, $H$) (left panel) and VR vs.\ VR($f$, $H$) (right panel), using the optimized parameters from \Tab {tab:trimming_dijet}.  The area of the untrimmed algorithm is roughly $\pi R_0^2$, as expected, while the trimmed jet's area is much smaller.}
\label{fig:catchmentArea}
}

In \Fig{fig:RwithFSRdefinedCells}, we argued that the overlap of ISR/MI with FSR was minimal, so even though the naive area of the jets employed in our analysis is quite large, there should not be significant sensitivity to the effects of ISR/MI/pileup that we set out to avoid.  We can quantify this statement using the catchment area of a jet~\cite{Cacciari:2008gn}, allowing us to directly measure the sensitivity of the trimmed jet to uniform diffuse contamination.  We find that while the jets we use in trimming start with large areas, after the jet trimming procedure is applied the  active area decreases dramatically, as shown in \Fig{fig:catchmentArea}.  In fact, the active area after trimming is even less than that of the untrimmed jet which began with a smaller radius.

One caution, however, is that the catchment area only captures the sensitivity to soft contamination.  Trimming cannot guard against a fluctuation of ISR/MI/pileup that yields a hard subjet above the $f_{\rm cut}$ threshold.  In some ways, trimming accentuates such fluctuations, since the contamination cannot be averaged over a larger jet area and subtracted statistically using, e.g.\ the methods of Ref.~\cite{Cacciari:2008gn}.  Therefore, more detailed studies are needed to really understand such systematic biases.  

\section{Conclusions}
\label{sec:conclusions}

In this paper, we have proposed jet trimming as a way to improve jet reconstruction by mitigating the spatial overlap 
between FSR and ISR/MI/pileup in hadronic collisions.  This technique actively removes sources of 
contamination by exploiting the difference in scale between the hard emissions of FSR and the relatively soft emissions from ISR/MI/pileup.   While prior efforts had been made along similar lines, those efforts focused on removing contamination from the jets of heavy boosted objects.  We have shown that light parton jets benefit from methods that emphasize relative subjet $p_T$ instead of the number of subjets.

We presented an explicit algorithm that implements jet trimming.  Our algorithm begins with seed jets constructed through 
any means (here we employ anti-$k_T$ and VR), which are then reclustered using an inclusive $k_T$ algorithm 
and trimmed according to a subjet $p_T$ cut set relative to some hard scale determined by the kinematics of the event.
In two different kinematic configurations, we find large improvements in reconstruction efficiency from using 
trimmed jets.  Moreover, unlike our previous VR algorithm~\cite{Krohn:2009zg}, this improvement  was obtained without a drastic increase in the catchment area of the jet.

Further study is necessary to understand how this jet trimming procedure would affect jet systematic errors in an 
actual experimental context.  For example, jet energy scale systematics already require a correction from ISR/MI/pileup
contamination, and exactly how a jet energy correction would be applied in the case of trimmed jets is unclear.  
However, by addressing ISR/MI/pileup contamination on a jet-by-jet basis, we expect that the systematic uncertainty 
associated with trimmed jets should not be any worse than for fixed-radius jets.  Moreover, it would be interesting to see whether the systematic shift in the invariant mass peak from accidentally throwing away FSR subjets could be fixed through a simple jet energy rescaling.

Finally, while the improvement in reconstruction from trimming is already quite helpful, it is nowhere near the in-principle improvement we saw in \Sec{sec:benefits}.  Perhaps further advances can be made through a better choice of the $\Lambda_{\rm hard}$ parameter or a different subjet finding procedure.  Whether any jet trimming algorithm can ever hope to approach the theoretical limit in ISR/MI/pileup rejection is an important open question, but the gains already seen in a simple trimming algorithm recommend its use at the LHC.

\acknowledgments{The authors would like to thank Jason Gallicchio, Gavin Salam, Matt Schwartz, and Peter Skands 
for useful discussions.  J.T. thanks the Miller Institute for Basic Research in Science for funding support.  L.-T. W. is supported by 
the National Science Foundation under grant PHY-0756966 and the Department of Energy under grant DE-FG02-90ER40542.  J.T. and L.-T. W.  thank the Aspen Center for Physics for their hospitality during the initial stages of this work.}

\bibliography{jetbib}
\bibliographystyle{jhep}
\end{document}